\documentclass[12pt]{article}
\usepackage[utf8]{inputenx}
\pdfoutput=1
\usepackage{amsmath}
\usepackage{amssymb}
\usepackage{amsthm}
\usepackage{bbm}

\usepackage[T1]{fontenc}
\usepackage{mathptmx}
\usepackage[varqu]{zi4}

\usepackage[left=1in,right=1in,top=1in,bottom=1in]{geometry}
\usepackage[singlespacing]{setspace}
\usepackage[inline]{enumitem}
\setlist{nosep}
\usepackage[format=hang]{caption}

\providecommand{\keywords}[1]{\textbf{\textit{Keywords}:} #1}

\usepackage[title]{appendix}

\usepackage[dvipsnames]{xcolor}
\usepackage{graphics}
\usepackage[outdir=./]{epstopdf}
\usepackage{booktabs}
\usepackage{mathtools}
\mathtoolsset{multlined-width=0.9\linewidth,multlined-pos=b}
\usepackage{authblk}
\usepackage{tabularx}

\newcounter{theorem}
\theoremstyle{definition}
\newtheorem{defn}[theorem]{Definition}

\theoremstyle{plain}

\theoremstyle{remark}

\newcommand{\F}{\mathcal{F}} 

\newcommand{\E}{\mathbb{E}}	
\newcommand{\R}{\mathbb{R}}	
\newcommand{\Z}{\mathbb{Z}}	
\newcommand{\N}{\mathbb{N}}	
\newcommand{\1}{\mathbbm{1}} 

\usepackage{xspace}

\newcommand{\U}{\mathbb{U}}

\newcommand{\I}{\mathbb{I}}

\usepackage[colorlinks=true,citecolor=blue,linkcolor=blue,pdftitle={A Pure-Jump Market-Making Model for High-Frequency Trading},pdfauthor={Baron Law and Frederi Viens}]{hyperref}	
\usepackage[numbers,sort&compress]{natbib}

\allowdisplaybreaks

\begin{document}

\title{Market Making under a Weakly Consistent \\ Limit Order Book Model}
\author[1]{Baron Law}
\author[2]{Frederi Viens}
\affil[1]{Agam Capital}
\affil[2]{Michigan State University}
\date{28 Jan, 2020}
\maketitle	
\begin{center}
	\textit{Published in High Frequency}
\end{center}
\begin{abstract}
We develop a new market-making model, from the ground up, which is tailored towards high-frequency trading under a limit order book (LOB), based on the well-known classification of order types in market microstructure. Our flexible framework allows arbitrary order volume, price jump, and bid-ask spread distributions as well as the use of market orders. It also honors the consistency of price movements upon arrivals of different order types. For example, it is apparent that prices should never go down on buy market orders. In addition, it respects the price-time priority of LOB. In contrast to the approach of regular control on diffusion as in the classical \citet{Ave2008} market-making framework, we exploit the techniques of optimal switching and impulse control on marked point processes, which have proven to be very effective in modeling the order-book features. The Hamilton-Jacobi-Bellman quasi-variational inequality (HJBQVI) associated with the control problem can be solved numerically via finite-difference method. We illustrate our optimal trading strategy with a full numerical analysis, calibrated to the order-book statistics of a popular Exchanged-Traded Fund (ETF). Our simulation shows that the profit of market-making can be severely overstated under LOBs with inconsistent price movements.\\~\\
\end{abstract}

\section{Introduction} \label{chp_intro}
Market makers in modern electronic order-driven exchanges provide liquidity to the market by posting limit buy and sell orders simultaneously on both sides of the limit order book (LOB). They earn the bid-ask spread in each round-trip buy-and-sell transaction in return for bearing the risks of adverse price movements, uncertain executions, and adverse selections \citep{Glo1985,Kyl1985}.

In the now classical setting of \citet{Ave2008}, which we will call AS framework or AS model thereafter, the authors assume that the mid price $S_t^m$ follows Brownian motion and the arrival of buy or sell market order, hitting a limit order at a distance of $d$ from the mid price, is an independent Poisson process with intensity $\lambda(d) = A\exp(-k d)$ where $A >0,k > 0$ are constants. A small market maker strives to maximize her risk-adjusted wealth at the end of a trading period by controlling her bid price $S_t^b$ and ask price $S_t^a$ at different times, subject to the dynamics of the mid price $S_t^m$, her cash holdings $B_t$, her inventory $Q_t$, and market order arrivals on the bid and ask sides $N_t^b, N_t^a$. The market-making problem can thus be cast as a stochastic optimal control as follows:
\begin{align}
\max_{S_t^b,S_t^a} &\E(U(B_T+Q_TS_T^m))\\
dS_t^m &= \sigma dW_t\\
dB_t &= S_t^adN_t^a - S_t^bdN_t^b\\
dQ_t &= dN_t^b - dN_t^a\\
\lambda_b &= A\exp(-k(S_t^m-S_t^b))\\
\lambda_a &= A\exp(-k(S_t^a-S_t^m))
\end{align}
where $W_t$ represents Brownian motion and $N_t^b,N_t^a$ denotes Possion processes independent from $W_t$ with intensity $\lambda_b, \lambda_a$ respectively. The quantity $\sigma > 0 $ represents instantaneous volatility and $U(\bullet)$ is a concave utility function.

The AS framework is adapted from \citet{Ho1981}, which is originally designed for dealers to trade in a quote-driven market (e.g. bonds, OTC derivatives), where they give bid and ask quotes\footnote{Here the quote means a price quotation to clients, but later we will use quote loosely to mean bid or ask price.} to potential clients via phone calls or nowadays Bloomberg terminals. \citeauthor{Ave2008} replace the assumption of a monopolistic market maker with an infinitesimally small one, so that the reference (mid) price is exogenous. In addition, they also substitute optimal limit orders for optimal price quotes  in order to trade in a LOB. This turns a game-theoretic model into a pure stochastic one, with which researchers in mathematical finance is much more comfortable, and thus their framework has became the foundation of many recent research papers in stochastic market-making models (see Table \ref{Tbl:MM})\footnote{Interested readers can also refer to Appendix \ref{appendx} for a short history of market-making models.}.

\begin{table}[ht!]
	\centering
	\caption{Making-Making Papers since 2008}
	\label{Tbl:MM}
	\begin{tabular}{llp{.7\linewidth}}
		\toprule
		Year	& Authors & Contribution  \\ 
		\midrule
		2012	& \citeauthor{Fod2012}  & solve the subsolution of the HJB PDE under a more general setting \\ 
		2013	& \citeauthor{Gue2013}  & efficient algorithm to solve the HJB PDE\\ 
		2013	& \citeauthor{Gui2013}  & allow market orders and arbitrary limit order volume, restrict bid/ask prices to best quotes or 1 tick better, mid price as jump diffusion \\ 
		2013	& \citeauthor{Fod2013a}  & extend to multiple assets \\ 
		2014 & \citeauthor{Car2014} & introduce dependence of price with order arrivals via the drift term\\
		2014 & \citeauthor{Nys2014} & introduce model uncertainty\\
		2015 & \citeauthor{Car2015} & compare different penalty functions in the optimization objective\\
		2015 & \citeauthor{Fod2015}  & mid-price as Markov renewal process correlated with order arrivals \\ 
		2017 & \citeauthor{Car2017} & introduce model uncertainty\\
		2017 & \citeauthor{Gue2017} & solve the HJB PDE under a more general setting with multiple assets\\
		2018 & \citeauthor{Car2018} & model the conditional intensity of order arrivals based on volume imbalance\\
		2018 & \citeauthor{Eva2018} & closed-form approximate solution of the HJB PDE under multi-asset environment\\
		\bottomrule
	\end{tabular} 
\end{table}

However, simply exchanging price quote for limit orders does not turn a quote-driven market-making model into a good order-driven one, as the AS framework does not address many important LOB features, which we are going to highlight in the following sections\footnote{See also \cite{Gue2017}.}.

\subsection{Price Consistency}
In the AS framework, price and order arrivals are assumed to be independent, so price can rise on a large sell market order, which is clearly impossible in real world LOB trading. Because market makers are often on the wrong side of the trade, due to the presence of informed traders (adverse selection), the absence of this crucial dependency structure often generates large phantom gain scenarios, leading to exaggeration of the average profit of market-making strategies. In our simulated backtest in Section \ref{InconBT}, this price inconsistency may overstate the market-making profit by more than 50\%. 

\subsection{Price-Time Priority}
Since nearly all LOBs now use the price-time\footnote{Highest execution priority goes to limit orders having better price, and then to those with earlier time-stamps.} priority, changing the price or quantity of a limit order means loss of execution priority; however, the AS framework assumes there is no cost in changing the bid and ask prices\footnote{Section 2.4 in \cite{Ave2008}: \emph{These limit orders $p^b$ and $p^a$ can be continuously updated at no cost.}}, as the model is originally designed for quote-driven market, which obviously incurs no cost in altering quotes to clients.

The optimal bid and ask prices at time $t$ in the AS model is expressed as a distance from the mid price at current time $t$, not the mid price at the time one posts the limit order, thus following the optimal bid and ask prices means continuously changing your limit orders. For example, according to the model, the current optimal bid and ask should be 3 ticks from the mid price and you place the limit orders as prescribed. When the mid price moves up 1 tick, your limit buy and sell orders are now at a position of 4 and 2 ticks from the mid price. According to the model, you should immediately cancel them and post new orders that is 3 ticks from the current mid price. If you follow the model, your orders will hardly get executed in a LOB as your orders are always at the bottom of the bid and ask queues due to your continuous update\footnote{In the backtest section of \cite{Gue2013}, the authors need to tweak the market environment and trading strategy in order  for the model to make sense under a LOB.}. 

The AS model makes perfect sense when it is used under a quote-driven market as per originally designed in \citet{Ho1981}. For example, when a client first calls, you give her a quote of say \$9.97/10.03. A minute later, she calls again, and you give her a quote of \$9.98/10.04 as your reference price has moved up \$0.01. Changing quotes in a quote-driven market in this example does not cost anything but it is not the case under a LOB because of the rule of price-time priority.

\subsection{Price Ticks}
In a LOB, prices are only allowed on a fixed price grid\footnote{For stocks with price > \$1, the tick size is \$0.01.}. As a result, a price is indeed a pure-jump process and it has two dimensions: namely jump times and jump magnitudes. A diffusion model can only approximate the magnitudes of the jumps but cannot describe the properties related to timing of the jumps such as jump clustering in high-frequency trading. 

In addition, the optimization may not be useful in some models under the constraint of price tick. For example, after spending numerous hours crunching the PDE in high precision, the optimal bid price from the model is \$10.0123456789. Nonetheless you cannot place a limit order with such a price in the LOB; you can either place an order with limit price \$10.01 or \$10.02. In Section \ref{exec_prob}, one may even find that in many cases you do not need to waste time solving the PDE as the only viable option is the best bid or best ask.

\subsection{Execution Probability}	\label{exec_prob}
A crucial component of the AS model is the rate function $\lambda(d)=A\exp(-k d)$, which directly affects the execution probability of limit orders in a given interval. In their model, price is continuous, so $d$ is a continuous variable. However, because of the discrete nature of the price grid in LOB, the rate function can only be a step function rather than a smooth curve. 

Moreover, when the limit order is more than one tick from the best quote in some liquid stocks, the execution probability is extremely small (e.g. less than 3\% for E-mini S\&P future \citep{Pom2011}\footnote{This execution probability is not the probability of a limit order posted to the second-best eventually gets executed after series of price movements. Rather, it is the probability that a limit order in the current second-best queue gets executed by a very large market order that walks up the book.}). As a result, the optimization problem becomes unnecessary, as in this case, the only reasonable action is to peg the limit orders to the best quotes.

\subsection{Order Size}	
For the sake of simplicity, the AS model assumes that all market and limit orders are of the same size. Such an assumption may mask the risk of overtrading by the market maker. 

In practice, market maker will not put all limit orders at one single pair of optimal bid and ask prices as suggested by the AS framework; instead they will place a plethora of limit orders at many price levels in order to continuously maintain her priority in the LOB, while orders are executed. 

Nonetheless, the arrival of one large market order may raise her inventory to an unacceptable level, and this kind of overtrading risk cannot be revealed when all orders have the same size.

\subsection{Paper Layout}
This paper is organized as follows: in Section 2, we define the notion of order-book consistency and then in Section 3, we fully describe our implementation of a weakly consistent LOB. Our novel market-making model is fully depicted in Section 4, and Section 5 illustrates some properties of our model with the numerical solution. Section 6 provides the result of a simulated backtest and Section 7 concludes.

\section{Consistency of Limit Order Book Model}
In a full LOB model, the only ingredients are limit and market orders. All the other quantities, namely bid price, ask price, bid-ask spread, and depth of limit order queues can be derived from the occurrences of limit and market orders. In a reduced form level-one LOB, however, one only observes the events which happen on the best bid and best ask; thus, such a model does not contain all the information required to derive the price dynamics. As a result, the prices in many market-making models are exogenous, and such prices are often inconsistent with the order-book transactions. 

Before we discuss specific examples, we first define what we mean by a consistent LOB model. In the following sections, we will use $(S_t^b, S_t^a,  S_t^m = (S_t^b+S_t^a)/2, S_t = S_t^a - S_t^b)$ to denote the bid price, ask price, mid price and bid-ask spread respectively, and the corresponding small letters $(s^b,s^a,s^m,s)$ will be used to express their current ``states'' values, at a given point in time.

\begin{defn}[Consistent Limit Order Book Model] Let $\tau_m^b, \tau_m^a, \tau_l^b, \tau_l^a, \tau_c^b, \tau_c^a$ denote the arrival times of any market sell, market buy, limit buy, limit sell, limit buy cancellation, and limit sell cancellation orders, and the corresponding volume and price (limit order only) be represented by $v$ and $\pi$. 

A limit order book model is called \emph{consistent} if it satisfies all of the followings.
	\begin{enumerate}
	\item Direction Consistency
		\begin{itemize}
		\item On the arrival of marketable\footnote{Marketable buy/sell order is either a market buy/sell order or a limit buy/sell order with limit price  greater/lower than or equal to the ask/bid price.} sell/buy order, the bid/ask price cannot move up/down while the ask/bid price can only stay unchanged: 
		\begin{equation}
		\mathbb{P} \left( \left\{S_{\tau_m^a}^a \ge S_{\tau_m^{a-}}^a \right\} \bigcap \left\{S_{\tau_m^a}^b = S_{\tau_m^{a-}}^b \right\} \right) = \mathbb{P} \left ( \left\{S_{\tau_m^b}^b \le S_{\tau_m^{b-}}^b \right\} \bigcap \left\{S_{\tau_m^b}^a = S_{\tau_m^{b-}}^a \right\} \right) =1
		\end{equation}
		\item On the arrival of limit sell/buy order with price falling inside the bid-ask spread, the ask/bid price can only move down/up while the bid/ask price can only stay unchanged. If the limit order is outside the bid-ask spread, both ask and bid prices remain unchanged:
		\begin{gather}
		A_1 = \left\{S_{\tau_l^a}^b = S_{{\tau_l^{a-}}}^b \right\} \bigcap \left\{S_{\tau_l^a}^a < S_{{\tau_l^{a-}}}^a \right\} \bigcap \left\{\pi_l^a < S_{{\tau_l^{a-}}}^a \right\}  \\
		A_2 = \left\{S_{\tau_l^a}^b = S_{{\tau_l^{a-}}}^b \right\}  \bigcap \left\{S_{\tau_l^a}^a = S_{{\tau_l^{a-}}}^a \right\}  \bigcap \left\{\pi_l^a \ge S_{{\tau_l^{a-}}}^a \right\}\\
		A_3 = \left\{S_{\tau_l^b}^a = S_{{\tau_l^{b-}}}^a \right\} \bigcap \left\{S_{\tau_l^b}^b > S_{{\tau_l^{b-}}}^b \right\} \bigcap \left\{\pi_l^b > S_{{\tau_l^{b-}}}^b \right\} \\
		A_4 = \left\{S_{\tau_l^b}^a = S_{{\tau_l^{b-}}}^a \right\}  \bigcap \left\{S_{\tau_l^b}^b = S_{{\tau_l^{b-}}}^b \right\}  \bigcap \left\{\pi_l^b \le S_{{\tau_l^{b-}}}^b \right\} \\
		\mathbb{P}\left( A_1  \bigcup A_2 \right) =\mathbb{P}\left( A_3  \bigcup A_4 \right) = 1
		\end{gather}
		\item On the arrival of limit sell/buy order cancellation at the best ask/bid, the ask/bid can only move up/down while the bid/ask can only stay unchanged. If the cancellation is outside of best quotes, both bid and ask prices remain unchanged:
		\begin{gather}
		B_1 = \left\{S_{\tau_c^a}^b = S_{{\tau_c^{a-}}}^b \right\} \bigcap \left\{S_{\tau_c^a}^a \ge S_{{\tau_c^{a-}}}^a \right\} \bigcap \left\{\pi_c^a = S_{{\tau_c^{a-}}}^a \right\}  \\
		B_2 = \left\{S_{\tau_c^a}^b = S_{{\tau_c^{a-}}}^b \right\}  \bigcap \left\{S_{\tau_c^a}^a = S_{{\tau_c^{a-}}}^a \right\}  \bigcap \left\{\pi_c^a > S_{{\tau_c^{a-}}}^a \right\}\\
		B_3 = \left\{S_{\tau_c^b}^a = S_{{\tau_c^{b-}}}^a \right\} \bigcap \left\{S_{\tau_c^b}^b \le S_{{\tau_c^{b-}}}^b \right\} \bigcap \left\{\pi_c^b = S_{{\tau_c^{b-}}}^b \right\} \\
		B_4 = \left\{S_{\tau_c^b}^a = S_{{\tau_c^{b-}}}^a \right\}  \bigcap \left\{S_{\tau_c^b}^b = S_{{\tau_c^{b-}}}^b \right\}  \bigcap \left\{\pi_c^b < S_{{\tau_c^{b-}}}^b \right\} \\
		\mathbb{P}\left( B_1  \bigcup B_2 \right) = \mathbb{P}\left( B_3  \bigcup B_4 \right) = 1
		\end{gather}
		\end{itemize}
	\item Timing Consistency - The bid/ask price moves only at the instants of orders arrivals/cancellations:
	\begin{equation}
	\mathbb{P}\left( \left\{ S_t^b = S_{t^-}^b \right\} \bigcap \Bigl\{ S_t^a = S_{t^-}^a \Bigr\} \middle | t \notin \Gamma \right) = 1
	\end{equation}
	where $\Gamma$ is the set of all stopping times of market and limit orders.
	\item Volume Consistency
	\begin{itemize}
		\item 	If the volume of the marketable buy/sell order is equal to or larger than the depth of the best ask/bid queue ($Q^a_t, Q^b_t$), the ask/bid price moves up/down; otherwise it stays unchanged:
		\begin{gather}
		\mathbb{P}\left( \left( \left\{S_{\tau_m^a}^a > S_{{\tau_m^{a-}}}^a \right\} \bigcap \left\{v_m^a \ge Q_{{\tau_m^{a-}}}^a\right\}\right)  \bigcup \left(\left\{S_{\tau_m^a}^a = S_{{\tau_m^{a-}}}^a \right\}  \bigcap \left\{v_m^a < Q_{{\tau_m^{a-}}}^a \right\}\right) \right) = 1\\
		\mathbb{P}\left( \left( \left\{S_{\tau_m^b}^b < S_{{\tau_m^{b-}}}^b \right\} \bigcap \left\{v_m^b \ge Q_{{\tau_m^{b-}}}^b\right\}\right)  \bigcup \left(\left\{S_{\tau_m^b}^b = S_{{\tau_m^{b-}}}^b \right\}  \bigcap \left\{v_m^b < Q_{{\tau_m^{b-}}}^b\right\}\right) \right) = 1
		\end{gather}
		\item 	If the volume of the limit buy/sell cancellation is equal to the depth of the best ask/bid queue ($Q^a_t, Q^b_t$), the ask/bid price moves up/down; otherwise it stays unchanged:
		\begin{gather}
		\mathbb{P}\left( \left( \left\{S_{\tau_c^a}^a > S_{{\tau_c^{a-}}}^a \right\} \bigcap \left\{v_c^a = Q_{{\tau_c^{a-}}}^a\right\}\right)  \bigcup \left(\left\{S_{\tau_c^a}^a = S_{{\tau_c^{a-}}}^a \right\}  \bigcap \left\{v_c^a < Q_{{\tau_c^{a-}}}^a \right\}\right) \right) = 1\\
		\mathbb{P}\left( \left( \left\{S_{\tau_c^b}^b < S_{{\tau_c^{b-}}}^b \right\} \bigcap \left\{v_c^b = Q_{{\tau_c^{b-}}}^b\right\}\right)  \bigcup \left(\left\{S_{\tau_c^b}^b = S_{{\tau_c^{b-}}}^b \right\}  \bigcap \left\{v_c^b < Q_{{\tau_c^{b-}}}^b\right\}\right) \right) = 1
		\end{gather}
	\end{itemize}

	\end{enumerate}
\end{defn}

When direction consistency is violated, the market makers' profit may be significantly exaggerated. For example, when the price plunges after a sequence of sell market order, the market maker, which have a net long inventory by taking the opposite sides of the trades, will suffer a major loss. Were the price to violate the direction consistency and go up, the market maker will instead enjoy a windfall profit. More generally, since the market maker is frequently on the wrong side of the trade, that is, she buys from market sell orders and sell from market buy orders, a LOB model violating direction consistency will likely overstate the market makers' profit (see Section \ref{InconBT} for simulation result).

We also require that price updates due to aggressive orders to be instantaneous to prevent phantom opportunities arising from stale prices, which would otherwise create a very profitable trading strategy\footnote{It is not an arbitrage in the classical sense, since there can be a major market sell order right after such a buy.}. For example, one may buy at the stale price right after a large buy market order and wait for the price to fully reflect the order book status, assuming the direction consistency will be observed eventually. The direction, together with timing consistency, ensure that the LOB model faithfully reflect the price risk only from the order book events.

Without volume consistency, the overstatement in a direction consistency violation may be exacerbated. In general, the average size of aggressive market orders\footnote{Aggressive order is an order that moves the price.} is larger than that of non-aggressive ones. Thus the loss due to adverse price movements from the aggressive market orders will be further understated when the volume distribution of aggressive orders are not modeled properly. 

Nevertheless, the volume consistency is more difficult to achieve in a LOB model as we need to keep track of the order book depth. We associate a label of \emph{weakly consistent} to a LOB if the model only complies with direction and timing consistency.

As mentioned, the conditions defined above for order book consistency are simply the direct consequences of normal order book operations, and thus any full LOB model, regardless of the distributional assumptions, will be fully compliant.

\subsection{Examples of inconsistent models}\label{chp_price}
Most of the existing market-making models do not model the full LOB: prices are often exogenous, leading to inconsistency of price movements, particularly when it is used in high-frequency trading.

For example, \citet{Ave2008} model the mid prices as independent Brownian motions.
\begin{equation} dS_t^m = \sigma dW_t \end{equation}
Such a setting may not reproduce even some simple stylized facts. For instance, when a buy market order arrives, since the mid price is an independent Brownian motion, half of the time it will go down, sometimes significantly. Such unrealistic scenarios may severely overstate the profit of a market-making strategy. As the price is a diffusion, it moves continuously even without any orders.

Observing these and other issues associated with totally independent prices and order arrivals, a few authors have attempted to incorporate some dependency structure between these two processes. \citet{Car2014} divide the buy and sell market orders into influential ($\overline{M}_t^+,\overline{M}_t^-$) and non-influential ($\widetilde{M}_t^+,\widetilde{M}_t^-$) with ($\overline{M}_t^+,\overline{M}_t^-$,$\widetilde{M}_t^+,\widetilde{M}_t^-$) being a multivariate Hawkes process. The mid price $S_t^m$ is a diffusion coupled with the market orders via an unobservable mean-reverting process $\alpha_t$ as follows:
\begin{gather}
dS_t^m = (\nu + \alpha_t)dt + \sigma dW_t\\
d\alpha_t = -\zeta \alpha_t dt + \sigma_\alpha dB_t + \varepsilon^+ d\overline{M}_t^+ - \varepsilon^- d\overline{M}_t^-
\end{gather}
where $W_t$ and $B_t$ are independent Brownian motions and $\nu \in \R, \zeta, \sigma, \sigma_\alpha, \varepsilon^+,\varepsilon^- 0$ are all strictly positive constants.

When an influential buy market order $\overline{M}_t^+$ arrives, $\alpha_t$ jumps up and the mid price $S_t^m$ will have a larger drift. Nonetheless, there is nothing to prevent the $W_t$ term from having an even larger negative change that results in an overall downward price movement. Since the mid price $S_t^m$ is continuous, it will not jump even with the arrival of influential market orders.

Another way to introduce the dependency structure is to model the mid price as pure jump process correlated with the order arrivals. \citet{Bac2014a} assume that the mid price has the form 
\begin{equation}S_t^m = S_0^m + \delta(N_t^+ - N_t^-) \end{equation}
Together with the buy and sell market order $M_t^+, M_t^-$, the quadruplet $(N_t^+,N_t^-,M_t^+,M_t^-)$ forms a multivariate Hawkes process \cite{Haw1971,Haw1971a,Law2016}. 

Although the prices and market orders are now correlated via the cross-excitation feature in the Hawkes process, there is still a non-zero probability that price goes down after a buy market order. In addition, a multivariate Hawkes process is by definition a \emph{simple} point process, meaning that price jumps and market orders occur at exactly the same time with probability zero.

In \citet{Fod2015}, the mid-price is modeled as
\begin{gather}
S_t = S_0 + \delta \sum_{T_n \le t} J_n
\end{gather}
where $\delta$ is the tick size and  $\{(T_n,J_n)\}$ is a Markov renewal process with $T_n \in \R_+$ and $J_n \in \{-1,1\}$ representing jump times and jump directions respectively. The market order is postulated as a marked point process $M(dt \times dz)$ where the mark $z_n$ indicates the side (buy/sell) of the market order. The stochastic intensity of $M$ depends on the elapse time since last price jump $T_m$ and the conditional distribution of $z_n$ depends on the direction $J_m$ of last price jump.\footnote{One may notice that the common believe of "order moves price" is reversed in this model and the authors acknowledge that his purpose is to reproduce the dependence between price and order rather than modeling causality.} 

It again has the same problem as \cite{Bac2014a} where price and market are correlated but the direction consistency can still be violated with a non-zero probability. Also, there is nothing to guarantee the price and arrival point processes to jump at the same time even though the price and order arrival processes are correlated.

In all of the above examples, either the volumes of all orders are assumed to be same, or the volume is not used directly to control the price jumps. 

In the last 10 years, academic research on market making (See Table \ref{Tbl:MM}) has been mostly focusing on extending and fine-tuning the \citet{Ave2008} framework, with little attention to its practicality in modern order-driven exchanges. This framework is well-suited to quote-driven markets like fixed income and OTC derivatives, but ill-suited to order-driven markets like equity and futures, owing to the presence of price-time priority. But ironically, most examples in the above quoted research are applying their results to high-frequency electronic trading platforms, which are order driven.\footnote{Similar comment was raised by \citet{Gue2017}.}

\section{A Weakly Consistent Level-One Reduced-Form LOB} \label{WeakLOB}
\subsection{The Model}
\begin{table}[!ht]
	\centering
	\caption{Order Classification}
	\label{order_class}
	\begin{tabular}{llcc}
		\toprule
		Type & Order Arrival Event & Bid Price & Ask Price \\
		\midrule
		1 & aggressive market buy & 0 & +  \\
		2 & aggressive market sell & - & 0 \\
		3 & aggressive limit buy & + & 0 \\
		4 & aggressive limit sell & 0 & -  \\
		5 & aggressive limit buy cancellation & - & 0 \\
		6 & aggressive limit sell cancellation & 0 & + \\
		\midrule
		7 & non-aggressive market buy & 0 & 0  \\
		8 & non-aggressive market sell & 0 & 0 \\
		9 & non-aggressive limit buy & 0  & 0 \\
		10 & non-aggressive limit sell & 0  & 0 \\
		11 & non-aggressive limit buy cancellation & 0  & 0 \\
		12 & non-aggressive limit sell cancellation  & 0 & 0 \\
		\bottomrule
	\end{tabular}\\
{\footnotesize +/-/0 indicates the corresponding price goes up/goes down/stays unchanged}
\end{table}

Following \citet{Bia1995} and \citet{Lar2007}, all orders\footnote{We ignore the exotic order types such as iceberg orders, retail price improvement orders, pegged orders, etc.} falling on the top of the limit order book can be classified into one of the twelve types in Table \ref{order_class} according to their categories (limit, market, cancellation), sides\footnote{Sell include both long sell and short sell.} (buy, sell) and aggressiveness. As in \cite{Lar2007}, aggressive orders are the ones which move the bid or ask price. To be more precise, an aggressive market order is one which completely depletes the best bid or ask queue, an aggressive limit order is one with limit price inside the bid-ask spread and an aggressive cancellation is one which cancels the last remaining order in the best bid or best ask queue.

Let $N(t)=(N_1(t),...,N_{12}(t))$ be the multivariate \emph{simple}\footnote{A point process is called \emph{simple} if it has at most one jump at each point of time almost surely. For simplicity, we assume no two types of orders can arrive at exactly the same time. Nonetheless, the probability that two orders arrive at the exact same instant is close to zero as the Nasdaq/CME timestamp is down to nanosecond.} point process of the number of orders in each type up to and including time $t$. $M^a(t),\ M^b(t)$ denote the number of \emph{buy} and \emph{sell} market orders and $S^b(t),\ S^a(t)$ represent the \emph{bid} and \emph{ask} price respectively. The tick size $\delta$ of the price grid is assumed to be fixed. 

The following straightforward but important relations are immediately observed.
\begin{align}
M^a(t) &= N_1(t) + N_7(t) \label{eq1.1} \\
M^b(t) &= N_2(t) + N_8(t) \label{eq1.2} \\
S^a(t) &= S^a(0) + (N_1(t) + N_6(t) - N_4(t))\delta \label{eq1.3}\\
S^b(t) &= S^b(0) + (N_3(t) - N_2(t) - N_5(t))\delta \label{eq1.4}
\end{align}

Eqautions \eqref{eq1.1} and \eqref{eq1.2} are simply the definition of $M^a(t)$ and $M^b(t)$. The ask price $S^a(t)$ is the initial ask price plus the number of aggressive orders which move the ask price up, minus the number of aggressive orders which move it down, multiplied by the tick size $\delta$. Equation \eqref{eq1.4} for the bid price follows from the same logic.

Through these remarkably simple equations \eqref{eq1.1}-\eqref{eq1.4}, one can observe the dependency of price and order arrivals via the common components $N_1$ and $N_2$. For instance, when there is an aggressive buy market order (type 1), both the buy market order point process $M^a(t)$ and ask price $S^a(t)$ will jump at the same time (co-jump), but they can also jump separately upon the arrivals of other order types. From these equations, one also recognizes that an ask price cannot go down with a buy market order (type 1 or 7).

However, the price jumps caused by aggressive orders can be larger than one tick and this is especially important for small tick stocks, where the limit orders are sparsely populated in the price grid. Therefore in addition to the random jump times $\tau _n \in \R_+ = (0,\infty)$, we add the random marks $\xi_n \in \N = \{0,1,2,..,\}$, which correspond to the jump sizes (in ticks) of the aggressive orders ($\xi_n=0$ for non-aggressive orders), and $v_n \in \R_+$, representing the volumes of the orders.

The multivariate marked point process\footnote{See \cite{Bre1981} for an excellent introduction to marked point process.} is now denoted as $N_i(dt \times dv \times d\xi)$ and its compensator will have the form $\lambda_i(t)\mu_i(t,dv \times d\xi)dt$ where $\lambda_i(t)$ is the intensity of the ground process $N_i(dt \times \R_+ \times \N)$ and $\mu_i(t,dv \times d\xi)$ is the conditional mark (volume and jump) distribution. For ease of exposition, we will use the notation $N_i(dt \times dv)=N_i(dt \times dv \times \N)$, $N_i(dt \times d\xi)=N_i(dt \times \R_+ \times d\xi)$ and $(N_i+N_j)(dt \times dv \times d\xi)=N_i(dt \times dv \times d\xi) + N_j(dt \times dv \times d\xi)$. The joint model of prices and market orders now becomes:
\begin{align}
M^a(dt \times dv) &= (N_1+N_7)(dt \times dv) \\
M^b(dt \times dv) &= (N_2+N_8)(dt \times dv)\\
S^a(dt) &=  \int_{\Z_+} \xi\delta (N_1+N_6-N_4)(dt \times d\xi) \label{price1}\\
S^b(dt) &=  \int_{\Z_+} \xi\delta (N_3-N_2-N_5)(dt \times d\xi) \label{price2}
\end{align}

The bid-ask spread $S(dt) = S^a(dt) - S^b(dt)$ in our model is given by
\begin{gather} 
S(dt) = \int_{\Z_+} \xi\delta (N_1 +  N_2 - N_3 - N_4+ N_5+ N_6)(dt \times d\xi)
\end{gather}
Because of the negative term, $N_3$ and $N_4$, one may concern that the bid-ask spread $S(t)$ may fall below one tick $\delta$. However, if we look closely at order types $3$ and $4$ (limit orders inside the spread), they can only happen when $S(t^-)>\delta$; therefore $S(t)$ will never shrink below $\delta$. 

The simplest way to enforce this constraint is to restrict the intensity of $N_3,N_4$. Based on the result that when the intensity of a point process is zero at time $t^-$, the probability that an event happens at time $t$ is zero \citep[T12, p.31]{Bre1981}, we impose the condition\footnote{We can also use the equivalent form $\lambda_i(t) = \lambda'_i(t) \1(S(t)>\delta)$, see \citet[T10, p.29]{Bre1981} for a proof.} that
\begin{align}
\lambda_3(t) &= \lambda'_3(t) \1(S(t^-)>\delta)\\
\lambda_4(t) &= \lambda'_4(t) \1(S(t^-)>\delta)
\end{align}
where $\lambda'_{3}(t)$ and $\lambda'_{4}(t)$ are any predictable non-negative stochastic processes.

It is easy to see that our model is direction- and timing-consistent, however since we do not keep track of the depths of best quotes, our model is not volume-consistent. 

\subsection{Intensity and Mark Distributions}
In general, the intensities of the order arrival processes can be any predictable non-negative stochastic process. In the simplest case, they are assumed to be constant, resulting in Poisson processes for these order arrivals. However, one may argue that, for example, $N_1(t)$ is simply a thinned process of the total market buy order with thinning probability $\mathbb{P}(v_t \ge Q_{t^-}^a|\F_{t^-})$ where $Q_{t^-}^a$ is the amount of limit sell orders resting in the best ask and $v_t$ is the incoming buy market order size. Therefore the intensity should be stochastic depending on the current shape of the limit order book.

However, including $Q_t^a$ in the model would significantly increase the dimension of the state variables needed for modeling, as we need to keep track of the order flow at all price levels, not just the best bid/ask \cite{Con2010}\footnote{For example, we need to keep track of orders falling at the second best, otherwise we would have no way to compute the queue depth of the new best quote after an aggressive market order.}. Moreover, the practices of quote stuffing\footnote{Rapid placement and cancellation of large amount of limit orders.} \cite{Egg2016}  and spoofing\footnote{Submission of limit orders to create an illusion of demand/supply imbalance.} \cite{Lee2013} render the usefulness of the full LOB questionable, especially beyond the best quotes. As a result, the benefit of a full LOB model may not justify the added nontrivial complexity and this may explain the emergence of reduced-form models, which focus only on the top of the book \cite{Con2013a,Ave2011}. 

Should the arrival intensity be assumed constant, it could be interpreted as the intensity based on the thinning probability in the equilibrium distribution of the order book. Since the market maker is supplying liquidity throughout the whole trading day, the average shape of the order book provides a reasonable approximation for our purpose.

Regarding order volume, \citet{Gop2000} observe that the sizes of market orders have power law tails (e.g. Pareto distribution) while \citet{Mas2001} find that log-normal distribution also fits the data reasonably well. We will use a log-normal distribution in our numerical example in Section \ref{solved_example} but our framework is compatible with any distributional assumption.

Any positive discrete distribution can be used to model the magnitude of the price jump and it can be conditional on the history of arrival times and volumes. For example, a very large order will cause the price to jump multiple ticks while a series of large market orders is more likely to cause the price to jump even more as the liquidity dries up\footnote{The speed at which the LOB reverts to its normal/equilibrium shape after a large order is called resiliency \cite{Lar2007}.}. However, for the sake of simplicity, we will assume volume and jump size are independent and stationary in the solved example in Section \ref{solved_example}.

\subsection{Parameter Estimation}

Since we assume the arrival intensities are constant, all the order types, except aggressive limit orders (type 3 and 4), follow Poisson processes. The Maximum Likelihood Estimator (MLE) of Poisson intensity is well-known to be
\begin{equation}
\hat{\lambda}_i = \frac{N_i(T)}{T} \quad \quad i \ne 3,4
\end{equation} 
For the aggressive limit order 3 and 4, the intensity is constrained to be zero while the bid-ask spread is one tick, so the MLE becomes
\begin{equation}
\hat{\lambda}_i = \frac{N_i(T)}{\int_0^T \1(S_t > \delta)dt} \quad \quad i=3,4
\end{equation} 

For the jump and volume distributions, one can either assume a parametric distribution with parameters estimated from MLE or simply use the empirical distribution. For example, given the sequence of observed jump sizes $\{\xi_1,...,\xi_N\}$ of some order type, the empirical distribution of $\xi$ is simply
\begin{equation}
f_\xi(k) = \sum_{j=1}^N \frac{\1(\xi_j=k)}{N}
\end{equation}
On a sparsely populated LOB where the range of observed jump sizes is very large, one can fit a parametric distribution (e.g. Negative Binomial) to $\{\xi_j\}$. If we assume the volume $v$ follows a lognormal distribution, that is log($v$) $\sim$ N($\mu$,$\sigma^2$), the MLE of $\mu$ and $\sigma^2$ is simply the sample mean and variance of log($v_j$)'s.

For the joint density of volume and jump size for aggressive order types, we express it as $f_{v,\xi}(v,\xi) = f_\xi(\xi)f_{v|\xi}(v|\xi)$ and $f_\xi$ can be estimated as just described using the empirical distribution. Since $\xi$ is discrete, we can easily estimate the conditional volume distribution given a particular jump size $\xi=k$, similar to the unconditional one.

\section{The Market-Making Model}\label{chp_mm}

\subsection{Trading Environment}

Our market maker can choose to post limit orders at both bid and ask or withdraw from one or both sides of the market. The operating regimes on bid and ask are indicated by $R^b_t, R^a_t \in \{0,1\}$ respectively. For instance, in the buy only (ask-off) regime ($R^b_t=1,R^a_t=0$), the market maker will only post limit buy orders at the best bid.

In addition, the market maker can issue a market order (impulse) of size $\zeta$ to adjust her inventory immediately, subject to the cost of crossing the bid-ask spread\footnote{The size of each market order is assumed to be small enough such that it does not walk up the LOB, and both the bid and ask queues are non-empty with probability one.}, exchange fee $\eta$, as well as a fixed overhead cost $c^i$.

In this version, our market maker will not consider aggressive limit orders (limit orders inside spread) as in \cite{Gui2013}, where a limit order inside the spread could have a permanent effective fill rate higher than that resting on the previous best quotes. In our model, the effect of switching from one regime to another is to switch on/off the arrival of \emph{our} market orders, which follow an prescribed point process dynamics. In our terminology, the gain, which we do not model here, is the temporary increase in participation rate $\rho$ due to higher order priority, rather than the arrival intensity of market orders due to our price improvement of one tick.

\subsection{Modeling Assumptions}

We assume that the market maker has only a small and pre-decided participation rate $\rho$ (e.g. say 10\%) among all the transactions in the market, so that her orders have negligible influence on the order flow. There are three alternatives regarding the interpretation of this participation rate $\rho$:
\begin{enumerate}
	\item For each market order of size $v$, our market maker will execute $\rho v$ shares; this implies that the orders of the market maker are infinitesimal small and distribute evenly in the queues. \label{rho1}
	\item For each market order, there is a probability $\rho$ that it will hit a limit order from our market maker. This assumption is more reasonable when the average market order size is small compared with that of limit order from our market maker. 
	\label{rho2}
	\item For each market order of size $v$, if $v < v^*$, where $v^*$ is a fixed constant, there is a probability $\rho$ that it will hit a limit order from our market maker. If $ v \ge v^*$, $\rho v$ shares will be executed against our market maker. \label{rho3}
\end{enumerate}
For large-tick stocks\footnote{Large-tick stock means the tick size is large relative to the price \cite{Day2015}.}, since the stock price is comparatively small and the average order size is big, option \ref{rho1} is reasonable. While for small tick stocks, for exactly the opposite reason, option \ref{rho2} seems more appropriate. For instance, the price of Berkshire Hathaway Inc. class A on 8 May, 2014 is \$190,010 and most market orders are of size one share, so it is not reasonable to assume that the market maker execute a fraction of each market order as in option \ref{rho1}. Option \ref{rho3} is more complicated but it can adapt to market orders of various sizes. In this article, we will use option \ref{rho1} for illustration. 

The market maker may achieve her target execution profile by continuously adjusting her limit orders in the LOB to be roughly the proportion $\rho$ of the queue length at each price level, but the detailed mechanism is outside the scope of this paper.

Since nearly all stock exchanges implement the so-called price-time priority, withdrawal from the market involves loss of priority of the current limit orders. To fully account for this, one would need to involve delay integral differential equations, which would further complicate the model. Instead, we simply penalize the switching of regime from $i$ to $j$ with  cost $c^b(i,j,t,s),c^a(i,j,t,s)$ on the bid and ask side respectively at time $t$ with spread $s$. We suggest the following simple formula for $c^b,c^a$, while acknowledging that more research is needed in this area.
\begin{align}
c^b(0,1,t,s) &= \alpha \bar{q}^b \rho (s/2 + \epsilon) \ \min \left(\frac{T-t}{\bar{q}^b/((\lambda_2 \bar{v}_2 + \lambda_8 \bar{v}_8))},1 \right) \\
c^a(0,1,t,s) &= \alpha \bar{q}^a \rho (s/2 + \epsilon) \ \min \left(\frac{T-t}{\bar{q}^a/((\lambda_1 \bar{v}_1 + \lambda_7 \bar{v}_7))},1 \right) \\
c^b(1,0,t,s)&=c^a(1,0,t,s)=0
\end{align}
where $\bar{v}_i$ is the mean volume for order of type $i$, $\bar{q}^b,\bar{q}^a$ are parameters representing the queue length ahead of our market maker in switching-off mode, and $\alpha$ is a constant discount factor.

Rationales for these formulas are as follows: canceling limit orders, in order to stop providing liquidity, does not cost the market maker anything, so we set $c^b(1,0)=c^a(1,0)=0$. However, when the market maker want to re-providing liquidity after pulling out from the market, she cannot do so until all the orders ahead of her are executed. As we will see later in the full model, we assume for simplicity that market maker will be able to capture market orders once the regime becomes 1; thus in fact the penalization cost $c^b,c^a$ is the amount to subtract from the overstated profit of $\bar{q} \rho (s/2 + \epsilon)$\footnote{In each round trip transaction, the market maker earns $s + 2 \epsilon$ per share and we attribute half of it to each side of the transaction.} due to the price-time priority, where $\bar{q}$ can be $\bar{q}^b$ or $\bar{q}^a$ depending on the side. However, one is not guaranteed to make money just by posting limit orders. The price may move away from the initial quotes and the market maker may even suffer a loss. That is why we introduce the discount factor $\alpha$ to take this into consideration. 

In addition, when market close is near, the overstated profit decreases as there is little time left for the market-making activity to run. The average time to consume all the bona fide limit orders on the bid side ahead of the market maker is about $\bar{q}^b/(\lambda_2 \bar{v}_2 + \lambda_8 \bar{v}_8)$. When the remaining time $T-t$ is less than that, we use the factor $(T-t)/(\bar{q}^b/(\lambda_2 \bar{v}_2 + \lambda_8 \bar{v}_8)$ to prorate the switching cost. In other words, when the time is near market close, it is less costly to switch off in order to minimize the final liquidation cost.

We would like to stress that $\bar{q}^b,\bar{q}^a$ is not the average depth of the book as the market maker does not need to cancel all her orders in order to temporarily leave the market. If this were the case, there would be a long delay before she can re-provide liquidity. To leave the market briefly, she should simply keep canceling sufficient high priority orders such that the remaining ones will not be executed with a probability of, say 90\%. For instance, if the 90 percentile of the sell market order size is 9000 shares, she just need to cancel about 1000 shares, given that her participation rate is 10\% and her orders are distributed uniformly in the queue. After, say, a sell order of size 900 shares hits the bid, she would cancel an additional 100 shares so that her orders will never appear in the top 9000 shares. Thus in this example, $\bar{q}^b = 9000$. One can further fine-tune the switching cost formula but the key takeaway is that a switching penalty should be less than the full order book depth.

The penalty for impulse (market order) is the exchange fee $\eta$ and the cost of crossing the bid-ask spread, which is already reflected in the holding $B_t$, together with an overhead cost $c^i$, which we model as the unexpected slippage cost incurred when the price moves away before the market maker can send out the market order.
\begin{equation}
c^i = \beta  \delta  \rho  \bar{v}_{\max} \label{Eq:impulse_cost}
\end{equation}
We assume there is a chance $\beta$ that the price moves away 1 tick as the market maker delays the submission of the market order for various reasons. The average size of her market order is about $\rho \bar{v}_{\max}$ where $\bar{v}_{\max}$ the maximum of average volume of type 1,2,7,8.\footnote{Using the actual impulse size in \eqref{Eq:impulse_cost} may seems to be better aligned with our rationale, but the impulse cost needs to be bounded away from 0, so that continuously sending out infinitesimally small impulses will never be optimal. Alternatively, the impulse cost could be regarded as a parameter to control the delay of impulses.} We emphasize that the trade-off between switching and impulse is quite sensitive to $c^b,c^a,c^i$ (see Section \ref{tradeoff}) and a thorough consideration is required to make the final result useful.

One limitation of our model is that the average execution price under aggressive market order cannot be computed faithfully as we only know the price at the best bid/ask. As an approximation, we will assume the average execution price is simply the best bid/ask, even though aggressive market orders, by definition, can walk up the LOB. As we will see in Table \ref{Tab:param} in Section \ref{order_book_ex} on our order book example, aggressive market orders are rare \cite{Pom2011}. In addition, our approximation errs on the conservative side as it always understates the market maker's profit.

\subsection{Market Making Optimal Control Problem}
The evolution of the market maker's cash holding $B_t$ and inventory $Q_t$ depends on the regime $(R^b_{t^-}, R^a_{t^-})$. For instance, when the market maker does not post any limit order $(R^b_{t^-}=R^a_{t^-}=0)$, the change in $B_t,Q_t$ will be zero. When $R^a_{t^-}=1$, for each buy market order, her inventory will be decrease by $\rho v$ where $v$ is the volume of the market order. The cash received will be the effective share quantity $\rho v$ multiplied by ask price $S^a_{t^-}$, plus the rebate $\varepsilon$\footnote{We do not include transaction tax, clearing fee, broker commission, etc. in this model but they can be incorporated very easily.}. The logic on the bid size is similar.

This setting, by which the market maker's limit orders are assumed to be distributed uniformly in the queue, affords a major simplification as we do not need to deal with a full-blown order-book model and keep track of the priorities of the market maker's orders in the queues. Within any given regime, the quantity of orders execution by our market maker is determined by the participation rate $\rho$, and the decision variables are only when to switch (limit order) regimes, when to place impulses (market orders), and how many shares to trade for the impulse\footnote{The direction of impulse order is trivial as the market maker should always act to reduce the magnitude of the inventory due to the associated penalty.}.

Consequently, the control $u$, which lies in some admissible set $\U_{ad}$, is a sequence of ordered quadruples $\{(\tau_n,r^b_n,r^a_n,\zeta_n)\}_{n \ge 1}$, where $\tau_n$ is the stopping time of the switching and/or impulse.  $r^b_n,r^a_n \in \{0,1\}$ are the new regimes on the bid and ask queues respectively and $\zeta_n \in \I \subset \R$ is the signed impulse strength (number of shares to buy (positive) or sell (negative)) in a compact set $\I$.\footnote{The set of impulse strength $\I$ may depend on the current inventory level and other state variables such that, after the impulse, the inventory is still within the domain. However, we will not highlight the dependence in the symbol for clarity.} $r^b_n, r^a_n, \zeta_n$ are all measurable with respect to $\F_{\tau_n}$. If $r_n=r_{n-1}$, it indicates no change of regime. If $\zeta_n = 0$, it means there is only switching but no market order. When $r_n \ne r_{n-1}$ and $\zeta_n \ne 0$, the market maker switches regime and issues market order at the same time. Since our switching and impulse costs do not depend on the regime, the order of switching and impulse does not matter.

The market-making optimal control problem is to maximize value function $V$ of expected total wealth (cash + inventory) at the end of the period $T$, minus the total cost on inventory penalty ($\theta \int Q_t^2dt$) with risk aversion $\theta$, switching ($c^b(\bullet),c^a(\bullet)$) and impulse ($c^i$), by choosing an optimal control $u=\{(\tau_n,r^b_n,r^a_n,\zeta_n)\}_{n\ge1}$, subject to the dynamics of bid and ask prices $S^b_t, S^a_t$, order arrivals $N_i(t)$, cash $B_t$ and inventory $Q_t$.

When a limit order from the market maker is executed, she will receive a per-share rebate $\varepsilon \ge 0$. Whereas a per-share exchange fee $\eta \ge 0$\footnote{We assume the fee structure is not inverted \cite{Bat2016}.}, in addition to the unfavorable price due to crossing of the spread, will be imposed when the market maker remove excess inventory with market order.

We now state the mathematical formulation of our market-making problem.
\begin{defn}[Market-Making Optimal Control Problem]
\begin{gather}
V(t_0,b,q,s^b,s^a,r^b,r^a) = \sup_{u \in \U_{ad}} J(t_0,b,q,s^b,s^a,r^b,r^a,u) \\
J(t_0,b,q,s^b,s^a,r^b,r^a,u) = \E \bigg\{ B_T +
(S^b_T-\eta)Q_T^+ - (S^a_T+\eta)Q_T^-  -\theta \int_{t_0}^T Q_t^2 dt \notag \\
-  \sum_{\tau_n \in (t_0,T]} \Big( c^b(r_{n-1}^b,r_n^b,\tau_n,S_{\tau_n}^a - S_{\tau_n}^b) + c^a(r_{n-1}^a,r_n^a,\tau_n,S_{\tau_n}^a - S_{\tau_n}^b) + c^i\1(\zeta_n \ne 0) \Big)  \notag \\
\bigg| B_{t_0}=b, Q_{t_0}=q, S^b_{t_0}=s^b, S^a_{t_0}=s^a, R_{t_0}^b=r^b, R_{t_0}^a=r^a \bigg\}\\
B_t =  b + \int_{(t_0,t] \times \R_+} \rho v R_{r^-}^a (S_{r^-}^a +\varepsilon) (N_1+N_7)(dr \times dv)  \notag \\
- \int_{(t_0,t] \times \R_+} \rho v R_{r^-}^b (S_{r^-}^b-\varepsilon) (N_2+N_8)(dr \times dv)
-  \sum_{\tau_n \in (t_0,t]} \Big( (S_{\tau_n^-}^a+\eta)\zeta_n^+ - (S_{\tau_n^-}^b - \eta)\zeta_n^- \Big) \label{Bt} \\
Q_t = q-\int_{(t_0,t] \times \R_+} \rho v R_{r^-}^a (N_1+N_7)(dr \times dv) 
 + \int_{(t_0,t] \times \R_+} \rho v R_{r^-}^b (N_2+N_8)(dr \times dv)  + \sum_{\tau_n \in (t_0,t]} \zeta_n \label{Qt} \\
S^b_t = s^b+ \int_{(t_0,t] \times \Z_+} \xi\delta (N_3-N_2-N_5)(dt \times d\xi)\\
S^a_t = s^a+ \int_{(t_0,t] \times \Z_+} \xi\delta (N_1+N_6-N_4)(dt \times d\xi)\\
R_t^b =r^b\1_{[t_0,\tau_1)}(t)+\sum_{n\ge1} r_n^b \1_{[\tau_n,\tau_{n+1})}(t)\\
R_t^a =r^a\1_{[t_0,\tau_1)}(t)+\sum_{n\ge1} r_n^a \1_{[\tau_n,\tau_{n+1})}(t)\\
c^b(0,1,t,s) = \alpha \bar{q}^b \rho (s/2 + \epsilon) \ \min \left(\frac{T-t}{\bar{q}^b/(\lambda_2 \bar{v}_2 + \lambda_8 \bar{v}_8)},1 \right) \\
c^a(0,1,t,s) = \alpha \bar{q}^a \rho (s/2 + \epsilon) \ \min \left(\frac{T-t}{\bar{q}^a/(\lambda_1 \bar{v}_1 + \lambda_7 \bar{v}_7)},1 \right) \\
c^b(1,0,t,s) =c^a(1,0,t,s)= 0\\
c^i = \beta  \delta  \rho  \bar{v}_{\max}
\end{gather}
where
\begin{gather}
x^+ = (|x|+x)/2, \ x^- = (|x|-x)/2
\end{gather}
\end{defn}

\subsection{Solving the Optimal Control Problem}

\citeauthor{Tan1993} \cite{Tan1993} show that the value function of a combined optimal switch and impulse control\footnote{The combined switching and impulse in \citeauthor{Tan1993}'s paper is slightly different from ours as the switching and impulse cannot happen at the time in their setting.} of a diffusion process is the unique viscosity of a system of variational inequalities. On the other hand, \citeauthor{Bis2010} \cite{Bis2010} prove that the value function of a optimal switching of a Levy process is the uniqueness viscosity solution of a system of nonlocal variational inequalities. 

By combining the arguments in \cite{Bis2010,Tan1993}, we conjecture that the value function $V(t,b,q,s^b,s^a,r^b,r^a)$ is the unique viscosity solution\footnote{As discussed in \cite{Con2005}, the value function $V$ may not be differentiable with respect to time $t$ due to the pure jump nature of the state dynamics, so a classical $C^1$ solution may not exist for the control problem.} of the following Hamilton-Jacobi-Bellman quasi-variational inequality (HJBQVI), which is now an integral differential equation\footnote{Please be aware that $q^+/q^-$ is the positive/negative part of $q$ while $q_+/q_-$ are defined in equations \eqref{q}.}. A rigorous proof of this result deserves a full paper of its own and we will leave it to interested researchers in stochastic control and viscosity solution. In this paper, we are satisfied with finding a numerical procedure which can provide us useful insights to tackle the market-making problem.
\begin{gather}
\min \Bigl\{ -\partial_t V(\bullet) -\mathcal{L}V(\bullet)+\theta q^2, V(\bullet)-\mathcal{M}V(\bullet) \Bigr\}=0 \\
V(T,b,q,s^b,s^a,r^b,r^a)=b + (s^b-\eta)q^+ - (s^a+\eta)q^-
\end{gather}
where
\begin{gather}
\mathcal{L}V(t,b,q,s^b,s^a,r^b,r^a)=  \notag \\
+ \int_{\R_+} \sum_{\xi=1}^{\infty} \Big( V(t,b_+,q_-,s^b,s^a_+,r^b,r^a)
- V(t,b,q,s^b,s^a,r^b,r^a) \Big) \lambda_1 f_1(v,\xi)dv \notag \\
+ \int_{\R_+} \sum_{\xi=1}^{\infty} \Big( V(t,b_-,q_+,s^b_-,s^a,r^b,r^a)
- V(t,b,q,s^b,s^a,r^b,r^a) \Big) \lambda_2 f_2(v,\xi)dv  \notag \\
+ \sum_{\xi=1}^{((s^a-s^b)/\delta) - 1} \Big( V(t,b,q,s^b_+,s^a,r^b,r^a) - V(t,b,q,s^b,s^a,r^b,r^a) \Big) \lambda_3  f_3(\xi) \notag \\
+ \sum_{\xi=1}^{((s^a-s^b)/\delta) - 1} \Big( V(t,b,q,s^b,s^a_-,r^b,r^a) - V(t,b,q,s^b,s^a,r^b,r^a) \Big) \lambda_4  f_4(\xi)\notag \\
+ \sum_{\xi=1}^{\infty} \Big( V(t,b,q,s^b_-,s^a,r^b,r^a) - V(t,b,q,s^b,s^a,r^b,r^a) \Big) \lambda_5 f_5(\xi)\notag \\
+ \sum_{\xi=1}^{\infty} \Big( V(t,b,q,s^b,s^a_+,r^b,r^a) - V(t,b,q,s^b,s^a,r^b,r^a) \Big) \lambda_6 f_6(\xi) \notag \\
+ \int_{\R_+} \Big( V(t,b_+,q_-,s^b,s^a,r^b,r^a)
 - V(t,b,q,s^b,s^a,r^b,r^a) \Big) \lambda_7 f_7(v)dv \notag \\
+ \int_{\R_+} \Big( V(t,b_-,q_+,s^b,s^a,r^b,r^a)
- V(t,b,q,s^b,s^a,r^b,r^a) \Big) \lambda_8 f_8(v)dv \label{InfGen}
\end{gather}
\begin{gather}
b_- = b-r^b (s^b -\varepsilon) \rho v, \
b_+ = b+r^a (s^a +\varepsilon) \rho v, \
q_- = q - r^a \rho v, \
q_+ = q + r^b \rho v \label{q}\\
s_-^b = s^b - \xi \delta, \
s_+^b = s^b + \xi \delta, \
s_-^a = s^a - \xi \delta, \
s_+^a = s^a + \xi \delta
\end{gather}

\begin{gather}
\mathcal{M}V(t,b,q,s^b,s^a,r^b,r^a)=
\max_{(\tilde{r}^b,\tilde{r}^a,\zeta) \in \{0,1\}^2 \times \I \backslash \{(r^a,r^b,0)\} } \Big\{ V(t, b - (s^a+\eta)\zeta^+ + (s^b-\eta)\zeta^-, \notag\\
q+\zeta,s^b,s^a,\tilde{r}^b,\tilde{r}^a)
-c^b(r^b,\tilde{r}^b,t,s^a-s^b) - c^a(r^a,\tilde{r}^a,t,s^a-s^b) - c^i \1(\zeta \ne 0) \Big\}
\end{gather}
where $f_1(v,\xi), f_2(v,\xi)$ are the joint probability density functions of the volume and jump distribution of aggressive market orders, $f_3(\xi),..., f_6(\xi)$ are the probability mass functions of the jump distribution of aggressive limit orders and cancellations and $f_7(v), f_8(v)$ are the density of the volume distribution of non-aggressive market orders. We assume $\int_\Omega \|x\| f_i(x)dx < \infty \  \forall i$. $\mathcal{L}$ is the infinitesimal generator of the state processes. $\mathcal{M}$ is the so-called the intervention operator, which maximizes the value function by switching regimes and/or issuing impulses, which ever action results in the highest value.


The expression for the infinitesimal generator \eqref{InfGen} looks daunting; however, in each part of the expression, it simply records the transaction changes on the arrival of different order types, no complicated mathematically theories are involved. For instance, when a type 1 (aggressive market buy) order arrives, the inventory of the market maker will be reduced by $r^a \rho v$ and the cash collected is $r^a (s^a +\varepsilon) \rho v$. Moreover, since it is an aggressive order, the ask price will jump by $\xi \delta$. The integral computes the average change over different volumes $v$ and jump sizes $\xi$ and finally, the result is multiplied by $\lambda_1$ to scale the effect by the arrival intensity.

\subsection{Ansatz}
Using the standard ansatz in the AS framework, we reduce the dimension of our state variables by two. In particular, the optimal control depends only on the bid-ask spread rather than the bid and ask prices and the cash level becomes irrelevant to the control problem.

Introducing new state variables for $s = s^a - s^b, \ s^m = (s^b+s^a)/2$ for the spread and mid price, and using the $(s,s^m)$-based ansatz
\begin{equation}
V(t,b,q,s^b,s^a,r^b,r^a) = b + s^m q  +\Phi(t,q,s,r^b,r^a)
\end{equation}
one can see that, after some simple but tedious algebra, $\Phi(t,q,s,r^b,r^a)$ satisfies the following HJBQVI.
\begin{gather}
\min \Bigl\{ -\partial_t \Phi(\bullet)-\mathcal{L}\Phi(\bullet)+\theta q^2, \Phi(\bullet)-\mathcal{M}\Phi(\bullet) \Bigr\}=0 \label{HJB} \\
\Phi(T,q,s,r^b,r^a)= -(s/2+\eta)|q|
\end{gather}
where
\begin{gather}
\mathcal{L}\Phi(t,q,s,r^b,r^a) = \notag \\
\int_{\R_+} \sum_{\xi=1}^{\infty} \Big(  \Phi(t,q_-, s_+, r^b,r^a) - \Phi(t,q,s,r^b,r^a) 
+ \xi\delta q_-/2  + r^a \rho v (s/2+\varepsilon) \Big) \lambda_1 f_1(v,\xi)dv  \notag \\
+  \int_{\R_+} \sum_{\xi=1}^{\infty} \Big( \Phi(t,q_+, s_+, r^b,r^a) - \Phi(t,q,s,r^b,r^a)
- \xi\delta q_+/2 + r^b \rho v (s/2+\varepsilon)   \Big) \lambda_2 f_2(v,\xi)dv  \notag \\
+ \sum_{\xi=1}^{(s/\delta) - 1}\Big( \Phi(t,q,s_-,r^b,r^a)-\Phi(t,q,s,r^b,r^a) + \xi\delta q/2  \Big) \lambda_{3}  f_3(\xi) \notag \\
+ \sum_{\xi=1}^{(s/\delta) - 1} \Big( \Phi(t,q,s_-,r^b,r^a)-\Phi(t,q,s,r^b,r^a)  -  \xi\delta q/2 \Big) \lambda_{4}  f_4(\xi) \notag \\
+ \sum_{\xi=1}^{\infty} \Big( \Phi(t,q,s_+,r^b,r^a)-\Phi(t,q,s,r^b,r^a) - \xi\delta q/2 \Big) \lambda_5 f_5(\xi) \notag \\
+ \sum_{\xi=1}^{\infty} \Big( \Phi(t,q,s_+,r^b,r^a)-\Phi(t,q,s,r^b,r^a) + \xi\delta q/2 \Big) \lambda_6 f_6(\xi) \notag \\
+ \int_{\R_+} \Big(  \Phi(t,q_-,s,r^b,r^a) - \Phi(t,q,s,r^b,r^a) 
+ r^a \rho v (s/2+\varepsilon)  \Big) \lambda_7 f_7(v)dv  \notag \\
+ \int_{\R_+} \Big( \Phi(t,q_+,s,r^b,r^a) - \Phi(t,q,s,r^b,r^a) 
+ r^b \rho v (s/2+\varepsilon)  \Big) \lambda_8 f_8(v) dv
\end{gather}
\begin{gather}
s_- = s - \xi \delta, \
s_+ = s + \xi \delta
\end{gather}
\begin{gather}
\mathcal{M}\Phi(t,q,s,r^b,r^a)=
\max_{(\tilde{r}^b,\tilde{r}^a,\zeta) \in \{0,1\}^2 \times \I \backslash \{(r^b,r^a,0)\} } \Big\{ \Phi(t,q+\zeta,s,\tilde{r}^b,\tilde{r}^a) \notag \\
-c^b(r^b,\tilde{r}^b,t,s) - c^a(r^a,\tilde{r}^a,t,s) - c^i \1(\zeta \ne 0) -(s/2+\eta)|\zeta| \Big\}
\end{gather}
and the optimal control is given by
\begin{gather}
\tau_k = \inf \Bigl\{ t>\tau_{k-1} \Big| \Phi(t,Q_{t^-},S_{t^-},R^b_{t^-},R^a_{t^-}) = \mathcal{M}\Phi(t,Q_{t^-},S_{t^-},R^b_{t^-},R^a_{t^-}) \Bigr\} \label{OptCtl1} \\
(r^b_k,r^a_k,\zeta_k) = \underset{(\tilde{r}^b,\tilde{r}^a,\zeta) \in \{0,1\}^2 \times \I \backslash \{(R^b_{\tau_k^-},R^a_{\tau_k^-},0)\}}{\mathrm{argmax}} \ \Bigl\{ \Phi(\tau_k,Q_{\tau_k^-}+\zeta,S_{\tau_k^-},\tilde{r}^b,\tilde{r}^a) \notag \\
-c^b(R^b_{\tau_k^-},\tilde{r}^b,\tau_k,S_{\tau_k^-}) - c^a(R^a_{\tau_k^-},\tilde{r}^a,\tau_k,S_{\tau_k^-}) - c^i \1(\zeta \ne 0) 
-(S_{\tau_k^-}/2+\eta)|\zeta|  \Bigr\} \label{OptCtl2}
\end{gather}
where we recognize the regime-switching and impulse stopping times as the times where the portion $\Phi$ of the value function ansatz, which depends on the regime variables, is indifferent to the intervention operator, while the choice of regimes and size of impulses maximizes the said $\Phi$ at the time of action, subject to new inventory, fee and cost structure. 

\subsection{Action Thresholds}
In a typical scenario, a market maker will start the day with no inventory and provide liquidity on both side of the LOB. If her inventory exceed a certain threshold, she may either cancel limit orders on one side or send a market order to reduce the inventory. We define $q^b_{\text{off}}$ (bid-off) to be the threshold over which it is optimal to stop providing liquidity on the bid and $q^b_{\text{imp}}$ (bid-impulse) to be the one over which to send market order. The minimum of bid-off and bid-impulse is denoted as $q^b_{\text{action}}$ (bid-action). The threshold under which the market maker should resume trading on the bid side is denoted as $q^b_{\text{on}}$ (bid-on). The quantities on the ask sides are defined similarly. 

\begin{defn}[Action Thresholds]
	\begin{align}
	q^b_{\text{off}}(t,s) &= \inf \Bigl\{q \in \R \Bigm| \Phi(t,q,s,0,1) - c^b(1,0,t,s) \ge \Phi(t,q,s,1,1)\Bigr\}\\
	q^b_{\text{imp}}(t,s) &= \min \Bigg\{ 
	\inf \Bigl\{q \ge q_{\text{off}}^b(t,s)  \Bigm| \Phi(t,q+\zeta,s,0,1) + \zeta (s/2 + \eta) - c^i \ge \Phi(t,q,s,0,1) \ \exists \zeta < 0 \Bigr\}, \notag \\
	& \qquad \inf \Bigl\{q \in \R \Bigm| \Phi(t,q+\zeta,s,1,1) + \zeta (s/2 + \eta) - c^i \ge \Phi(t,q,s,1,1) \ \exists \zeta < 0 \Bigr\} 
	\Bigg\}\\
	q^b_{\text{action}}(t,s) &= \min \left\{q^b_{\text{off}}(t,s),q^b_{\text{imp}}(t,s) \right\} \\
	q^b_{\text{on}}(t,s) &= \sup \Bigl\{q \le q_{\text{off}}^b(t,s) \Bigm| \Phi(t,q,s,1,1) - c^b(0,1,t,s) \ge \Phi(t,q,s,0,1)\Bigr\}\\
	q^a_{\text{off}}(t,s) &= \sup \Bigl\{q \in \R \Bigm| \Phi(t,q,s,1,0) - c^a(1,0,t,s) \ge \Phi(t,q,s,1,1)\Bigr\}\\
	q^a_{\text{imp}}(t,s) &= \max \Bigg\{ 
	\sup \Bigl\{q \le q_{\text{off}}^a(t,s)  \Bigm| \Phi(t,q+\zeta,s,1,0) - \zeta (s/2 + \eta) - c^i \ge \Phi(t,q,s,1,0) \ \exists \zeta > 0 \Bigr\}, \notag \\
	& \qquad \sup \Bigl\{q \in \R \Bigm| \Phi(t,q+\zeta,s,1,1) - \zeta (s/2 + \eta) - c^i \ge \Phi(t,q,s,1,1) \ \exists \zeta > 0 \Bigr\} 
	\Bigg\}\\
	q^a_{\text{action}}(t,s) &= \max \left\{q^a_{\text{off}}(t,s),q^a_{\text{imp}}(t,s) \right\} \\
	q^a_{\text{on}}(t,s) &= \inf \Bigl\{q  \ge  q_{\text{off}}^a(t,s) \Bigm| \Phi(t,q,s,1,1) - c^b(0,1,t,s) \ge \Phi(t,q,s,1,0)\Bigr\}
	\end{align}
\end{defn}

If $q^b_{\text{imp}} < q^b_{\text{off}}$, it is optimal to use market order to eliminate excess inventory to avoid the risk of adverse price movement due to uncertain execution. On the other hand, if $q^b_{\text{off}} < q^b_{\text{imp}}$, it is beneficial to wait for the offsetting order, rather than paying the bid-ask spread and exchange fee.

We will illustrate the relationship of these thresholds with other model parameters in Section \ref{solved_example}.

\subsection{Numerical Scheme} \label{NumericScheme}

We present here a numerical method using finite difference similar to the so-called penalty scheme\cite{Ben1984,Bel2009}. The HJBQVI \eqref{HJB} can be approximated by the following representation when the parameter $\gamma > 0$ is sufficiently small.\footnote{It can be verified easily that for any $c>0$, $\min(x,y)=0$ iff $\min(x,cy)=0$ and $\min(z+x,z+y)=z+\min(x,y)$.}.
\begin{align}
&\min \Bigl\{ -\partial_t \Phi -\mathcal{L}\Phi+\theta q^2, 
\Phi-\mathcal{M}\Phi - \gamma(\partial_t \Phi +\mathcal{L}\Phi-\theta q^2) \Bigr\}=0\\
\iff &\min \Bigl\{ -\partial_t \Phi -\mathcal{L}\Phi+\theta q^2,
(\Phi-\mathcal{M}\Phi)/\gamma - (\partial_t \Phi +\mathcal{L}\Phi-\theta q^2) \Bigr\}=0\\
\iff &-\partial_t \Phi -\mathcal{L}\Phi+\theta q^2 + \min \Bigl\{ 0, (\Phi-\mathcal{M}\Phi)/\gamma \Bigr\}=0 \\
\iff &-\partial_t \Phi -\mathcal{L}\Phi+\theta q^2 - (\mathcal{M}\Phi - \Phi)^+/\gamma =0 
\end{align}

Replace the $t$-derivative by backward difference, we have a discretized version $\Phi^h$
\begin{equation}
-(\Phi^h(t_{n+1}) - \Phi^h(t_n))/h_t - \mathcal{L}^h\Phi^h(t_{n+1},\bullet) + \theta q^2 - (\mathcal{M}^h\Phi^h(t_{n+1},\bullet) - \Phi^h(t_{n+1},\bullet))^+/\gamma =0
\end{equation}
Rearranging the equation, we arrive at our explicit numerical scheme.
\begin{gather}
\Phi^h(T,q,s,r^b,r^a)= -(s/2+\eta)|q| \\
\Phi^h(t_n,\bullet) = \Phi^h(t_{n+1},\bullet) + h_t \Bigl( \mathcal{L}^h\Phi^h(t_{n+1},\bullet) - \theta q^2  \Bigr) + \frac{h_t}{\gamma} \Bigl(\mathcal{M}^h\Phi^h(t_{n+1},\bullet) - \Phi^h(t_{n+1},\bullet) \Bigr)^+   \label{Eq:scheme}
\end{gather}

$\Phi^h$, where $h=(h_t,h_q,h_I,h_M)$, will be computed on a finite grid of time and space (spread, inventory, bid and ask). The step size for the backward difference is $h_t$ whereas the inventory is divided into $N_q$ intervals of size $h_q$, and $\Phi^h$ is computed up to maximum value of $N_s$ for the spread state variable. The integrals inside the $\mathcal{L}^h$ operator can be evaluated numerically using any quadrature technique (e.g. trapezoidal rule) with step size $h_I$. For values of $\Phi^h$ between grid points in the numerical integration, one can assign a value using the method of nearest-neighbor interpolation. The maximum in the $\mathcal{M}^h$ operator can be found by exhaustive search on a compact subset $\I^h \subseteq \I$ with grid size $h_M$. 

After solving for $\Phi^h$ on the grid, the optimal control can be obtained naturally from
\begin{gather}
\tau_k = \min \Bigl\{ t_n>\tau_{k-1} \big| \Phi^h(t_n,Q_{t_n},S_{t_n},R^b_{t_n},R^a_{t_n}) \le \mathcal{M}^h\Phi^h(t_{n},Q_{t_n},S_{t_n},R^b_{t_n},R^a_{t_n}) \Bigr\} \\
(r^b_k,r^a_k,\zeta_k) = \underset{(\tilde{r}^b,\tilde{r}^a,\zeta) \in \{0,1\}^2 \times \I^h \backslash \{(R^b_{\tau_k},R^a_{\tau_k},0)\}}{\mathrm{argmax}} \ \Bigl\{ \Phi^h(\tau_k,Q_{\tau_k}+\zeta,S_{\tau_k},\tilde{r}^b,\tilde{r}^a) \notag \\
-c^b(R^b_{\tau_k},\tilde{r}^b,\tau_k,S_{\tau_k}) - c^a(R^a_{\tau_k},\tilde{r}^a,\tau_k,S_{\tau_k}) - c^i \1(\zeta \ne 0) 
-(S_{\tau_k}/2+\eta)|\zeta|  \Bigr\} \label{Eq:impulse}
\end{gather}

\section{Numerical Illustration}
\subsection{An Order Book Example} \label{order_book_ex}
A summary of the order book statistics of QQQ (PowerShares QQQ ETF) on May 8, 2014 12pm - 3pm\footnote{It is well-known that market has different behavior during the opening and closing period \cite{Bia1995}, so we only use the data between 12-3pm.} is shown in Table \ref{Tab:QQQ}. \emph{However, we would like to stress that the numbers are based on the order flow in the Nasdaq LOB only. Since Nasdaq  executed only 20.8\% (by shares)\footnote{https://www.nasdaqtrader.com/trader.aspx?ID=marketsharedaily} of QQQ on May 2014 among all US stock exchanges, the number of aggressive orders is likely to be overestimated\footnote{NYSE TAQ contains consolidated trades and quotes from all US exchanges, but the timestamps of the trades and quotes are not synchronized \cite{Lee1991} and subjected to significant delay \cite{Din2014}.}. The figures here are meant for illustration only.} 

QQQ is actively traded in Nasdaq; its bid-ask spread is one tick most of time and the depths of the best quotes are reasonably healthy.
\begin{table}[!ht]
	\centering
	\caption{Order Book Statistics of QQQ on May 8, 2014, 12pm-3pm (Nasdaq LOB only)}
	\label{Tab:QQQ}
	\begin{tabular}{lr}
		\toprule
		Feature & Value \\
		\midrule
		Average Mid Price (\$)	& 87.0040  \\ 
		Average Bid-Ask Spread (tick)	& 1.0775 \\ 
		Average Best Bid Depth (shares)	&  18,650 \\ 
		Average Best Ask Depth (shares)	&  19,796 \\ 
		\bottomrule
	\end{tabular} 
\end{table}

\begin{table}[!ht]
	\centering
	\caption{Type Level Statistics of QQQ on May 8, 2014 12-3pm (Nasdaq LOB only).\\
		$\lambda$ is the arrival intensity, $\bar{v}$ and $\bar{\xi}$ are the mean of volume (in shares) and jump size (in ticks)}
	\label{Tab:param}
	\begin{tabular}{llrrrrr}
		\toprule
		Type & Description                            & Count   & \% Count & $\lambda$ (/s) & $\bar{v}$ & $\bar{\xi}$ \\
		\midrule
		1    & aggressive market buy                  & 730     & 0.10\%   & 0.0676      & 766         & 1.0000           \\
		2    & aggressive market sell                 & 669     & 0.09\%   & 0.0619      & 1,174         & 1.0000           \\
		3    & aggressive limit buy                   & 1,499   & 0.20\%   & 0.2140      & 1,134         & 1.0000           \\
		4    & aggressive limit sell                  & 1,625   & 0.22\%   & 0.2320      & 1,056         & 1.0000           \\
		5    & aggressive limit buy cancellation      & 937     & 0.13\%   & 0.0868      & 706         & 1.0000           \\
		6    & aggressive limit sell cancellation     & 789     & 0.11\%   & 0.0731      & 637         & 1.0000           \\
		7    & non-aggressive market buy              & 1,118   & 0.15\%   & 0.1035      & 814         & NA  \\
		8    & non-aggressive market sell             & 1,124   & 0.15\%   & 0.1041      & 724         & NA           \\
		9    & non-aggressive limit buy               & 176,421 & 23.89\%  & 16.3353     & 1,052         & NA           \\
		10   & non-aggressive limit sell              & 184,271 & 24.95\%  & 17.0621     & 1,061         & NA           \\
		11   & non-aggressive limit buy cancellation  & 182,498 & 24.71\%  & 16.8980     & 959         & NA          \\
		12   & non-aggressive limit sell cancellation & 186,943 & 25.31\%  & 17.3095     & 946         & NA 		\\        
		\bottomrule
	\end{tabular}
\end{table}

The type level statistics are presented in Table \ref{Tab:param}. Because of the rapid placements and cancellations of limit orders, potentially including e.g. quote stuffing \cite{Egg2016} or spoofing \cite{Lee2013}, 98.83\% of the order flow belongs to types 8 thru 12. Fortunately, thanks to our assumption of uniform distribution of market maker's limit orders, these types do not factor into our market-making model. This is one of the reasons why we do not intend to pursue a full order-book model since the dominating activities of limit-order placements and cancellations will only increase the complexity without adding any value to explain the decision process of a bona-fide market maker \cite{Ber2017}. 

The relatively large aggressive limit order arrival rates $\lambda_3, \lambda_4$, together with the almost sure jump size of one tick, contribute to the tight spread of the QQQ LOB. The sum of intensities for all market order types is 0.3371. This means that there is a trade executed on the Nasdaq LOB around every 3 seconds on average. The average sizes for aggressive and non-aggressive market buy orders are about the same but the average size of aggressive market sell orders is 60\% larger than that of non-aggressive ones. It is expected that one needs large volumes to move prices, but in the current trading environment, nearly all institutional investors use algorithmic trading to trade large blocks, and the algorithmic engine will typically divide the blocks into small pieces to hide its intention.

\subsection{A Solved Example} \label{solved_example}

We solve the HJBQVI \eqref{HJB} numerically for a 5-minute trading session ($T=300s$) using finite difference as describe in Section \ref{NumericScheme} with the parameters depicted in Table \ref{Tab:ex_param}, which are realistic since they resemble those we just discussed for QQQ. However, we make the parameters on the buy and sell sides symmetric so as not to embed any alpha view in the model (see Section \ref{OrdImb}). In addition, we ensure $\lambda_3,\lambda_4$ are sufficiently large so that the spread will not diverge under the long simulation horizon in Section \ref{bt}. 

For the mark distribution, we simply assume volumes $v_i$ are independent from the jump sizes $\xi_i$ and $v_1$ and $v_2$ follow the lognormal$(6.5,1.35^2)$ law while $v_7$ and $v_8$ follow the lognormal$(6,1.35^2)$ law. For the jump size (in ticks) $\xi_1,\xi_2$, we use the Bernoulli law $\mathbb{P}(\xi=1) = 1 - \mathbb{P}(\xi=2) = 0.95$ and  $\xi_3,\ldots,\xi_6$ are 1 tick almost surely. We use Simpson's rule to compute the integrals in the $\mathcal{L}^h$ operator with step size $h_I$ of 100 shares on a bounded interval $(0, \exp(\mu_i + 2 \sigma_i))$ for each order type. Since $\rho$ is 0.1 and the step size of $\Phi(q)$ is 10 shares, no interpolation is required to get the value of $\Phi(q_-), \Phi(q_+)$.
\begin{table}[!ht]
	\centering
	\caption{Base Case Parameters}
	\label{Tab:ex_param}
	\begin{tabular}{m{49pt}rm{49pt}rm{49pt}rm{49pt}r}
		\toprule
		Market Parameter & Value & Exchange Parameter & Value & Model Parameter & Value & Discretization Parameter & Value \\ \midrule
		$\lambda_1,\lambda_2$ & 0.05 & $\delta$ & 0.01 & $\theta$ & 1e-7 & T & 300s \\
		$\lambda_3,\lambda_4$ & 0.25 & $\epsilon$ & 0.002 & $\rho$ & 0.1 & $h_t$ & 0.1s \\
		$\lambda_4,\lambda_5$ & 0.075 & $\eta$ & 0.003 & $\bar{q}^b$ & 3750 & $h_q$ & 10 shares \\
		$\lambda_7,\lambda_8$ & 0.1 &  &  & $\bar{q}^a$ & 3750 & $N_q$ & 2001 \\
		&  &  &  & $\alpha$ & 0.3 & $N_s$ & 8 \\
		& &  &  &$\beta$ & 0.1 & $h_I$ & 100 shares \\
		& &  &  & & & $h_M$ & 10 shares \\
		& &  &  &  & & $\gamma$ & 0.1 \\
		\bottomrule
	\end{tabular}
\end{table}

Unless stated otherwise, all the charts in this section show the values when spread = 1.
\subsubsection{Trade-off between Switching and Impulse} \label{tradeoff}
\begin{figure}[ht!]
	\centering
	\includegraphics[width=0.5\linewidth]{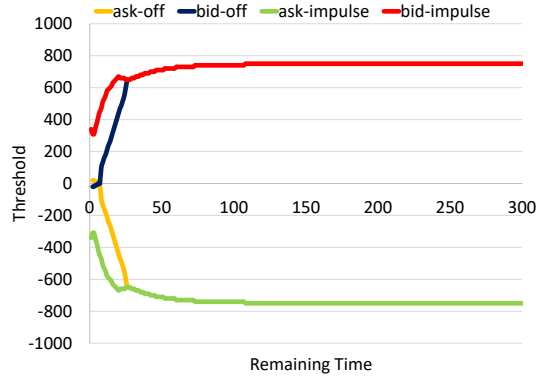}
	\caption{Off and Impulse Thresholds vs Remaining Time}
	\label{fig:off}
\end{figure}

Figure \ref{fig:off} plots the graph of ask-off, bid-off, ask-impulse and bid-impulse thresholds vs remaining time to close $(T-t)$ when the spread is one tick. The chart is symmetric with respect to the bid and ask due to the symmetric parameters, so we will only show the bid side thereafter. 

The bid-off goes to infinity at 26 seconds in remaining time, meaning that the only optimal action is to send market order when the inventory exceeds the bid-impulse limit. In other words, throughout the whole trading session, it is not optimal to switch off the bid until shortly before close. 

This is a bit surprising since, traditionally, one assumes that a market maker is a liquidity provider, so she should rarely, if at all, use market orders. This is true in quote-driven markets as there is simply no market order available. However, under an order-driven market, market orders in fact provide a very effective means to control inventory risk.

The reason is that when one uses market order, one does not really pay the spread, one just does not earn it. Suppose the market maker has just reached the impulse threshold of 750 shares and then she accepts a sell market order to buy an additional 100 shares at \emph{bid}. She can immediately issue a market sell order to get rid of this unwanted 100 shares at the same \emph{bid} price, provided that she is quick enough so that the bid price has not moved. The cost of unwinding this trade is simply the exchange fee $\eta$ - rebate $\epsilon$, which is only 0.001 per share (0.1 tick). 

If instead she cancels $\bar{q}^b \rho$ shares in the bid queue, the lost opportunity cost under our framework when spread equals 1 is 0.3*3750*0.1*(0.01/2+0.002)=\$0.7875. Taking into account the impulse overhead $c^i=\$0.1$ and assuming average order size of 1000, this is equivalent to the cost of roughly 4 market orders. That means if the first market buy order comes right after 4 consecutive market sell orders, the two approaches roughly cost the same but if the first market buy order comes earlier, the impulse approach will cost less. Since our parameters are symmetric, the probabilities of buy and sell order arrivals are equal; thus the probability of seeing at least 4 consecutive sell orders is about 6\%. In addition, if the price goes down 1 tick, the market maker will suffer a loss of 0.01*100=\$1, by not using a market order.

Nonetheless, switching may become effective when the decay factor of the switching cost kicks in; at that time the market maker should start unwinding the inventory to avoid paying the exchange fee at the final liquidation. Besides, you will see in Section \ref{bs-spread} that when the spread is sufficiently large, it is justified to switch off before sending impulse in order to earn the spread.

There is also a catch when using market order for inventory control. When adverse selection occurs, the incoming market order triggering the breach of threshold may be aggressive, so the market-maker may suffer loss as the last incoming market order has already depleted the best quote and she can no longer unwind the excess inventory at cost.

The above illustration is highly simplified, as various types of buy and sell market orders have different intensities, volume, and jump characteristics. However, after considering the distributional assumptions, fees, rebates and risk aversion, the model result still tells us that market orders are indeed a very efficient tool for the market maker to control inventory risk. 

The impulse threshold reaches its equilibrium value of 750 shares very quickly at 108 seconds remaining. This result coincides with that of \cite{Gue2013,Gui2013}. We have solved the HJBQVI up to 1 hour and, though not reported here for the sake of conciseness, we have noted that the thresholds remain the same.

\subsubsection{Switching and Impulse Cost}
\begin{figure}[ht!]
	\begin{minipage}{0.5\linewidth}
		\centering
		\includegraphics[width=1\linewidth]{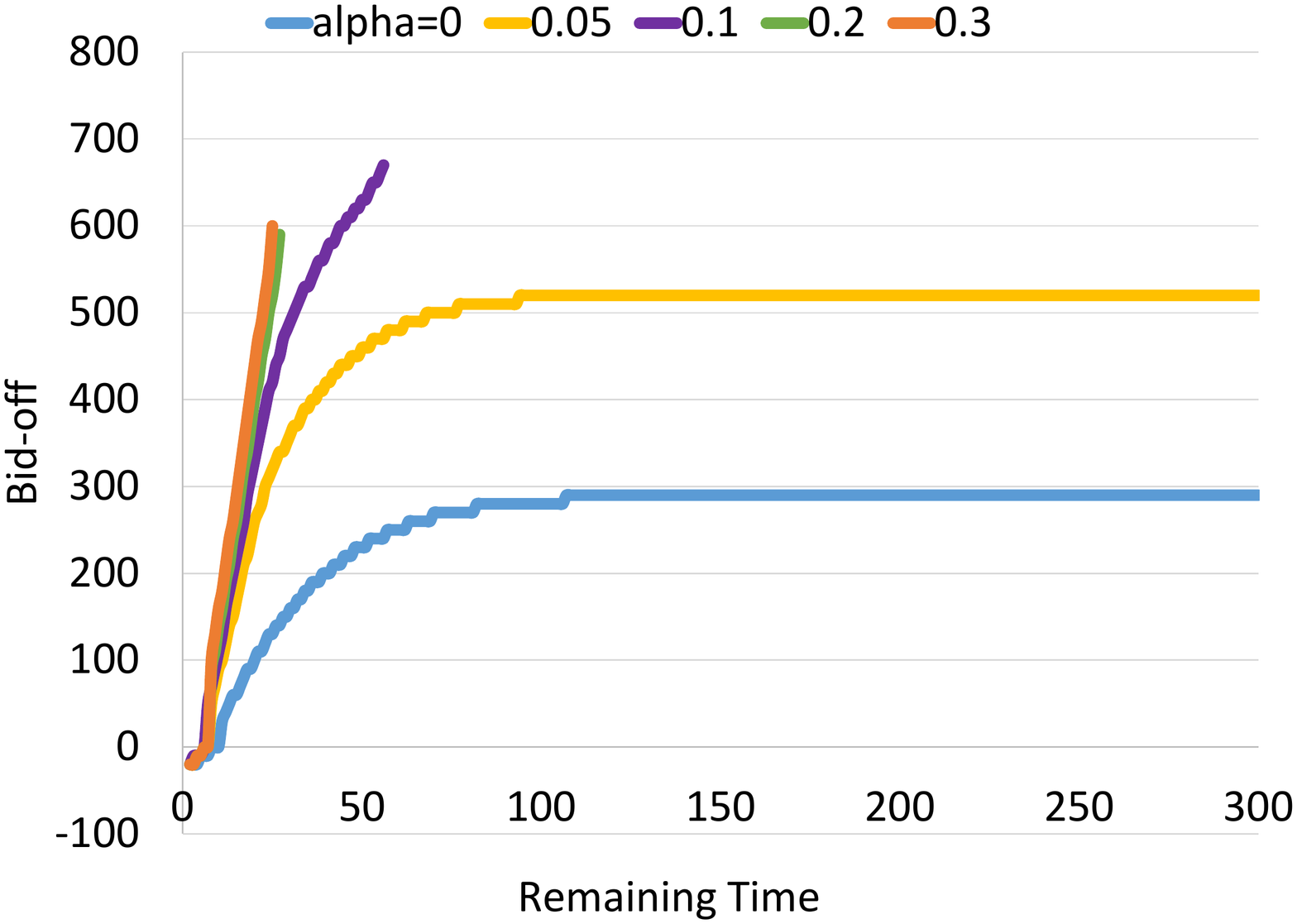}
		\caption{Bid-Off with Alpha}
		\label{fig:alpha}
	\end{minipage}
	\begin{minipage}{0.5\linewidth}
		\centering
		\includegraphics[width=1\linewidth]{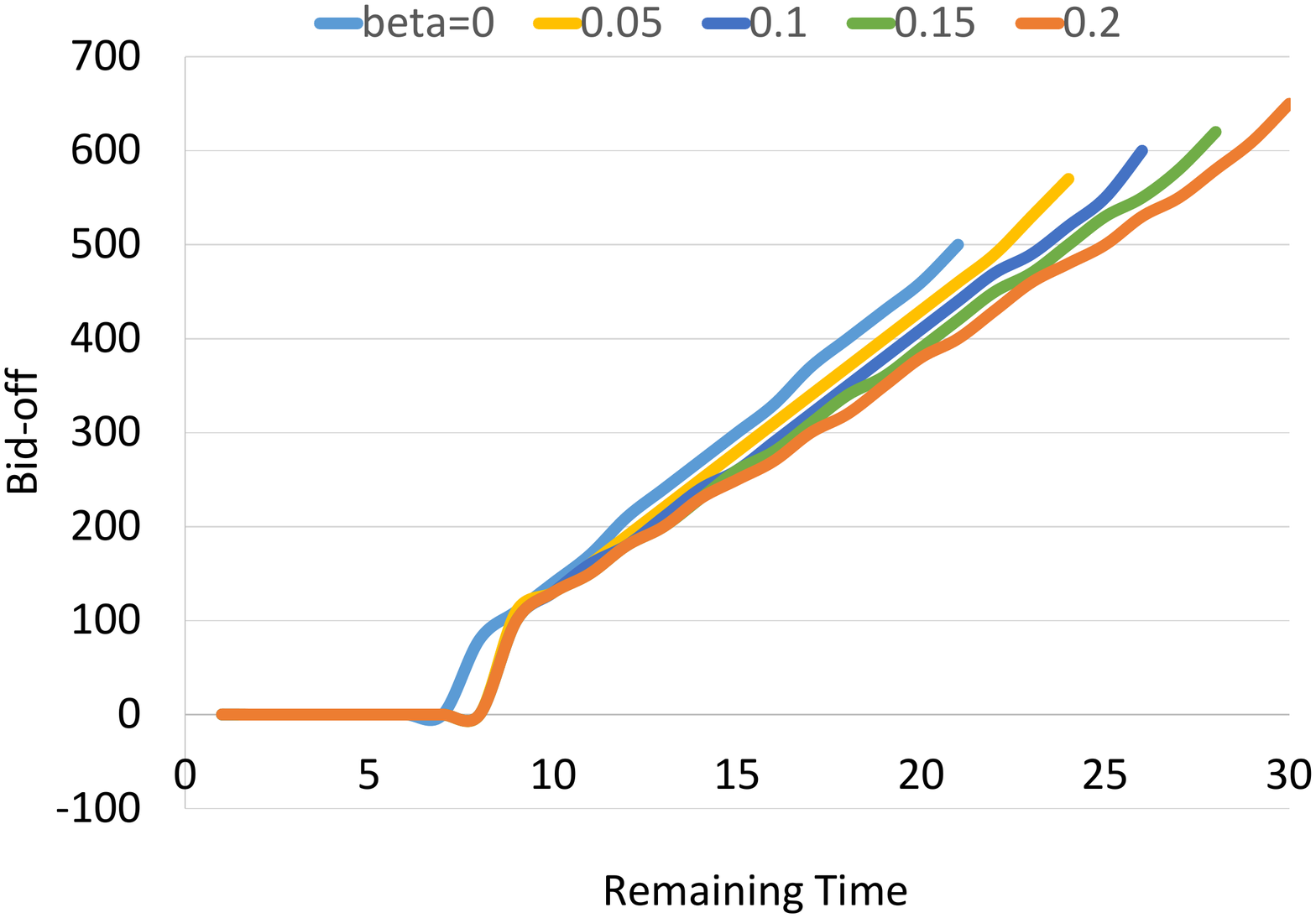}
		\caption{Bid-Off with Beta}
		\label{fig:beta}
	\end{minipage}
\end{figure}

The appeal of switching depends largely on the switching cost $c^b,c^a$, which is modulated by the discount factor $\alpha$. We plot the graphs of bid-off against time for various choice of $\alpha$ in Figure \ref{fig:alpha}.

We can see that when $\alpha \le 0.05$, switching become more attractive, compared with market order impulses, but one should remember that our interpretation of $\alpha$ is the chance of bid or ask price stays at the same level or moves favorably to the market maker, til the other leg of a round-trip market-making transaction is executed. It is a judgment call to decide whether such a low level of $\alpha$ is reasonable.

On the other hand, switching is also preferred when impulses are costly. Transaction fee is fixed by the exchange, but under our framework there is a fixed cost $c^i$, which is modulated by another discount factor $\beta$. Figure \ref{fig:beta} shows a high $\beta$ lowers the threshold, as the slippage renders market orders less effective, leading to a more conservative optimal strategy. On the other hand, if the market maker is concerned about the slippage due to unexpected adverse selection, she can also increase $\beta$ accordingly.

\subsubsection{Optimal Impulse Size}
\begin{figure}[ht!]
	\centering
	\includegraphics[width=0.5\linewidth]{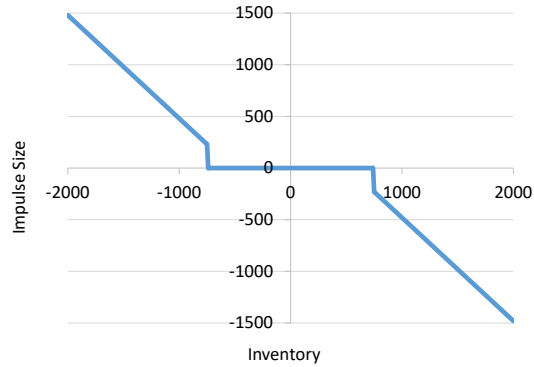}
	\caption{Optimal Impulse Size when Spread=1, remaining time = 300s}
	\label{fig:impulse}
\end{figure}
Figure \ref{fig:impulse} shows the optimal size of the market order at $T-t = 300$s against the inventory level, when the impulse threshold is exceeded. Because of the fixed impulse cost $c^i$, the impulse size starts at 230, but not 0,  when the inventory reaches 750, and then it increases linearly with inventory level.

The linearity of impulse size vs inventory makes the optimal decision rule much simpler to use. In this example, the optimal impulse size is simply inventory minus 520, which we called impulse \emph{anchor}, and it is roughly the impulse threshold when $c^i=0$. The impulse anchor reaches the equilibrium very quickly; the optimal impulse graph is exactly the same whether the remaining time is 300s or 3600s. Because of this, we do not need to maintain a large table describing the optimal impulse size at each time, spread, and inventory level, we just need to store the bid and ask impulse thresholds and anchors for each spread.

\subsubsection{Bid-Ask Spread} \label{bs-spread}
\begin{figure}[ht!]
	\centering
	\includegraphics[width=0.5\linewidth]{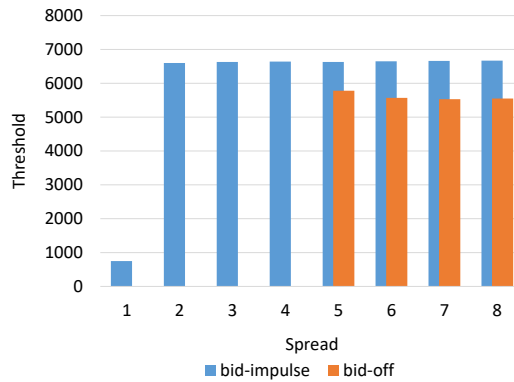}
	\caption{Bid-Impulse and Bid-off with Spread}
	\label{fig:spread}
\end{figure}

The Bid-ask spread is crucial to the market maker's profit; thus we expect the optimal action will change according to the prevailing spread and Figure \ref{fig:spread}, which shows the bid-off and bid-impulse thresholds under different spreads, confirms our intuition.

The impulse limit increases when the spread widens but still, impulse is preferred over switching when the spread is relatively low. However when the spread is large (spread $\ge$ 5), the market-making business become so lucrative that the greed starts to trump the fear. In this case, the market maker will switch off the bid before sending sell market order, hoping that she can eventually earn the spread.

Our result is similar to that of \cite{Gui2013} in the sense that the impulse threshold increases with spread. However in \cite{Gui2013}, the market maker always switches off first before sending out market orders as there is no switching cost in that paper.

In Figure \ref{fig:spread}, the impulse threshold jumps when the spread is greater than one tick. It is because a two-tick spread is much more profitable than a one-tick spread. Under a one-tick spread, if the price move one tick against the market maker after one side of the transaction is executed, her profit will be just the rebate (2/5 tick). However, under a two-tick spread, she earns one tick plus rebate (7/5 tick), which is 3.5 times the one-spread case.

Another reason is that in our example the type 1,2 orders jump one tick with probability 0.95 and type 3-6 orders always jump one tick. Were the jump size distribution be more dispersed, the change in threshold from one to two ticks will be less prominent. In addition, the relatively large $\lambda_3,\lambda_4$ ensure that the spread is one tick most of the time, so this state transition should rarely occur in our example.
	
\subsubsection{Order Imbalance} \label{OrdImb}
\begin{figure}[ht!]
	\begin{minipage}{0.5\linewidth}
		\centering
		\includegraphics[width=1\linewidth]{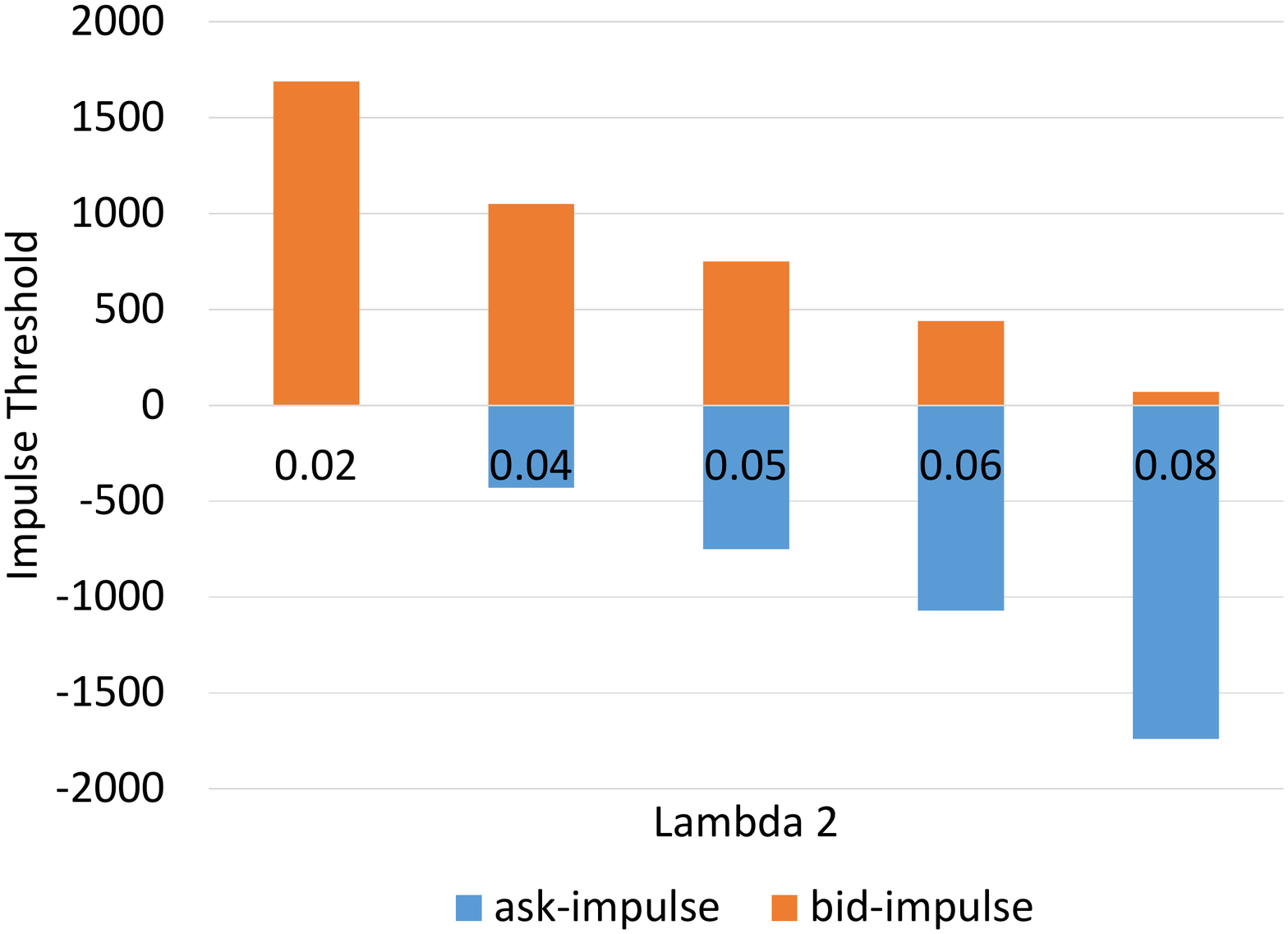}
		\caption{Impulse Thresholds with Different $\lambda_2$}
		\label{fig:lambda2}
	\end{minipage}
	\begin{minipage}{0.5\linewidth}
		\centering
		\includegraphics[width=1\linewidth]{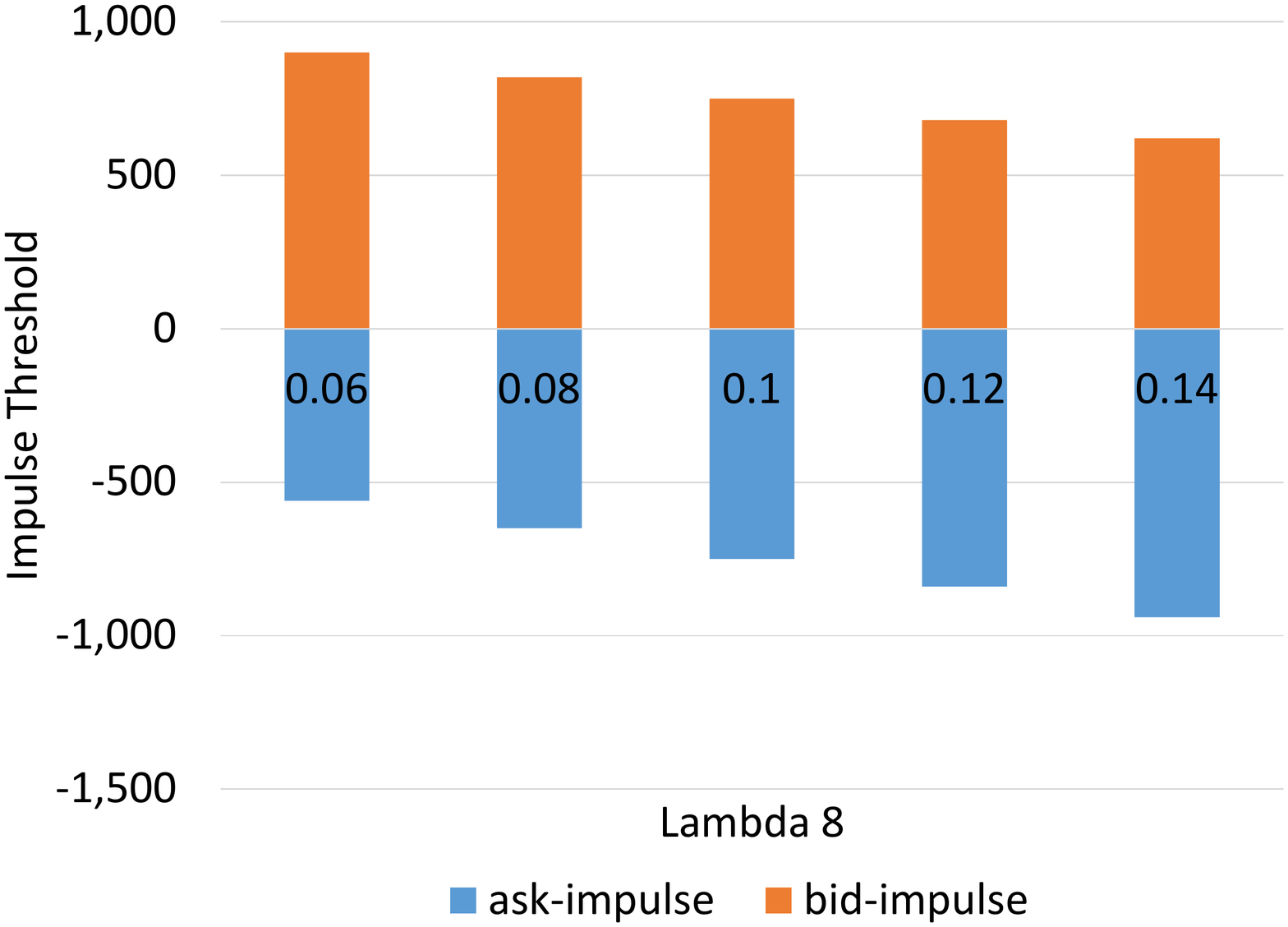}
		\caption{Impulse Thresholds with Different $\lambda_8$}
		\label{fig:lambda8}
	\end{minipage}
	\begin{minipage}{0.5\linewidth}
		\centering
		\includegraphics[width=1\linewidth]{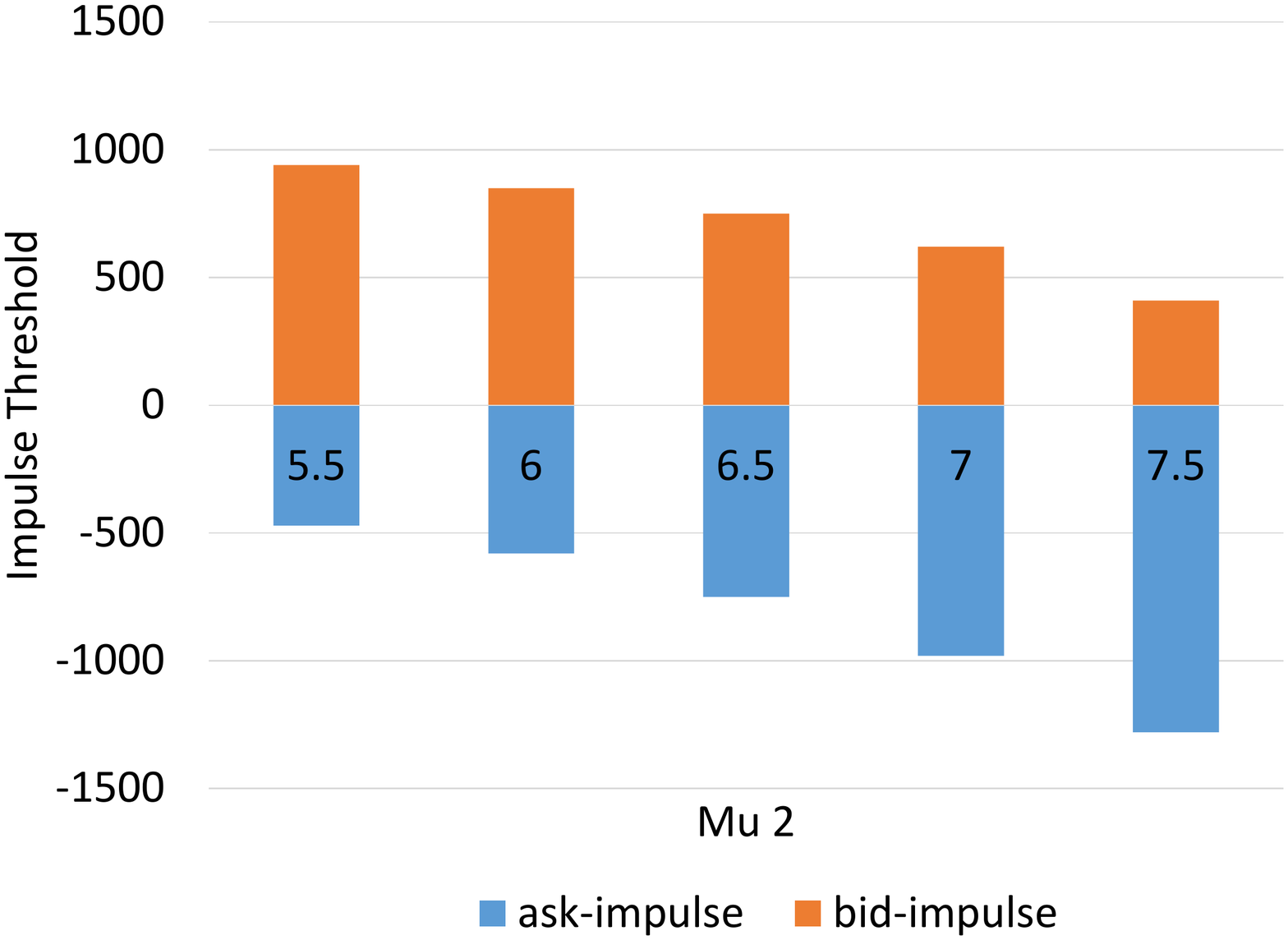}
		\caption{Impulse Thresholds with Different $\mu_2$}
		\label{fig:mu2}
	\end{minipage}
	\begin{minipage}{0.5\linewidth}
		\centering
		\includegraphics[width=1\linewidth]{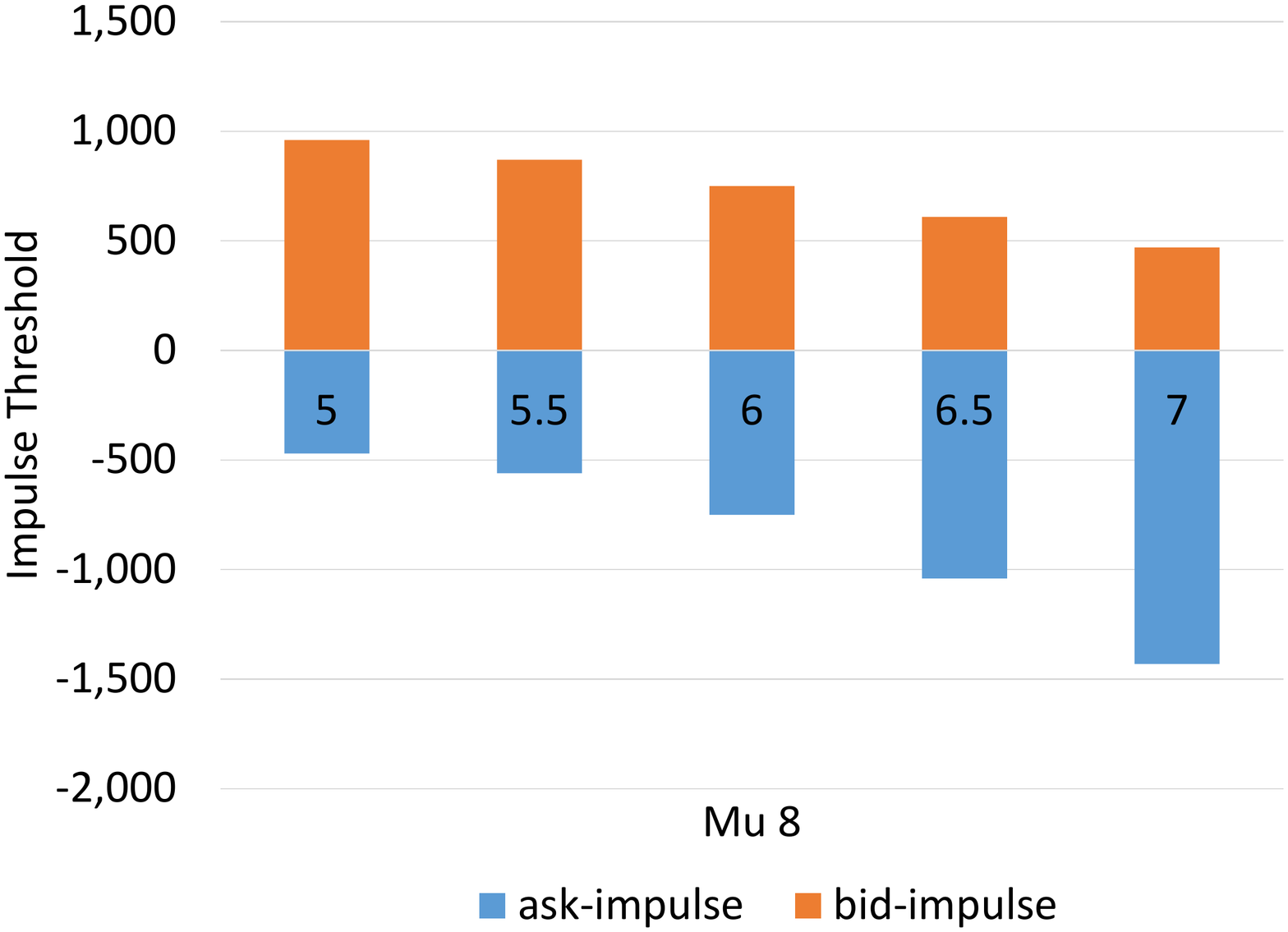}
		\caption{Impulse Thresholds with Different $\mu_8$}
		\label{fig:mu8}
	\end{minipage}
\end{figure}

With symmetric distributions between buy and sell orders, the thresholds on the bid and ask sides are symmetric. However when the order distributions become skewed, the optimal action under this implied alpha view is to scale back the market-making activity\footnote{This agrees with empirical evidence \cite{Ana2016}.}, as this is the way in which market makers protect themselves from adverse selection \citep{Glo1985,Kyl1985}.

Figure \ref{fig:lambda2} shows the impulse thresholds with different intensities $\lambda_2$ for aggressive sell market orders. When $\lambda_2$ = 0.08 (large selling pressure), the market maker should maintain her inventory below 70 using market orders, effectively not providing any liquidity on the bid side. The case is similar when $\lambda_2 = 0.02$ (large buying interest) and it is not optimal for the market to provide liquidity on the ask side as she will suffer huge loss due to adverse price movements.

The effect of $\lambda_8$ (Figure \ref{fig:lambda8}) has a similar but much smaller effect on the thresholds, as type 8 orders are not aggressive; it just takes longer to liquidate the inventory in such market, but the market orders themselves will not introduce adverse price movements against the market maker. Figure \ref{fig:mu2}, \ref{fig:mu8} show a similar trend for mean volume and we do not repeat the argument again.

\section{Simulated Backtest} 
\subsection{Backtest under a Consistency LOB} \label{bt}

In this section, we run a simulation to test the performance of two market-making strategies under our weakly consistent LOB defined in Section \ref{WeakLOB}, with the configuration parameters the same as in Table \ref{Tab:ex_param}. 

Simply put, twelve types of orders, modeled as multivariate marked point processes\footnote{See \cite{Law2016} for methods to simulate marked point processes.}, with marks being volume and jump size, will arrive at the top of the LOB, where $\rho=10\%$ of the limit orders are from our market maker and they are uniformly distributed in the queues. The arrivals of different type of orders will trigger the changes in prices and/or bid-ask spread according to Table \ref{order_class}. The market maker's cash and inventory will evolve according to equations (\ref{Bt},\ref{Qt}). Our market maker can cancel her limit orders on one or both sides of the queues and/or send market orders to control her inventory level if necessary. 

The two strategies are described below:
\begin{enumerate}
	\item Unconstrained Trading - the market maker continuously provide liquidity on both sides of the market with no risk limit.
	\item Optimal Control - the market maker will act according to the optimal control (\ref{OptCtl1}, \ref{OptCtl2}) with the admissible set of the impulse $\I = [-M,M]$ where $M \gg 0$. In this simulation, we solve for the optimal control only up to 300 seconds from close, and the decision rule at that point is extended to the earlier session.
\end{enumerate}

In order to make the simulation more realistic regarding the price-time priority in a real LOB, the market maker's limit orders will not be executed after switching on until $\bar{q}^b/\bar{q}^a$ shares of limited orders ahead of them have been eliminated. On the other hand, even when the market maker is in switch-off mode, if the order size is greater than $\bar{q}^b/\bar{q}^a$, she will still execute the excess portion, subject to her usual participation rate $\rho$. This is the price to pay as she wants to maintain priority in future executions by canceling only limit orders at the top of the book.

We will not deduct the artificial cost of switching $c^a,c^b$, impulse $c^i$ and inventory penalization $\int \theta Q_t^2 dt$ in the simulation; only real cost such as exchange fees and rebates are included. The initial setting are: $S_0^b = 100, S_0^a = 100.01, Q_0 = 0, B_0=0, R_0^b=R_0^a=1$. The length of each trading session is 6.5 hours and one millions iterations were executed. The performance of the two strategies are shown in Table \ref{Tbl:bt1}.
\begin{table}[ht!]
	\centering
	\caption{Simulated Backtest on the Weakly Consistent LOB (N=1E6, T=6.5 hours)}
	\label{Tbl:bt1}
		\begin{tabular}{lrr}
			\toprule
			& Unconstrained Trading &   Optimal Control\\
			\midrule
			Mean & 8,224 & 6,471 \\
			Std \mbox{Deviation} & 10,258 & 516 \\
			Skewness & -0.44 & 0.10  \\
			Kurtosis & 7.81 & 3.19 \\
			IR = Mean/SD & 0.80 & 12.55  \\
			 \bottomrule
		\end{tabular}%
\end{table}

The mean profit under the optimal control strategy is slightly smaller than that under the unconstrained strategy, but the standard deviation and kurtosis are much smaller while the skewness changes from negative to positive. Also, the large reduction in standard deviation relative to the mean profit leads to a much better reward-to-risk or \emph{information ratio} (IR). Though we did not calibrate the risk aversion parameter $\theta$ to maximize the IR, the IR, together with standard deviation, skewness, and kurtosis, is significantly improved as a result of our strategy's sound risk management decision.

\subsection{Backtest under Inconsistency LOBs} \label{InconBT}

In order to demonstrate the importance of order book consistency, we also simulate the performance of the same two strategies under hypothetical order books which suffer from inconsistency as follows: all the order flows are exactly same as in Section \ref{bt}, except for the direction of price movements, which now depend on an independent random variable. Suppose that a type 1 (market buy) order arrives, the pseudo code for the weakly consistent model in Section \ref{bt} is
\begin{verbatim}
ask_price = ask_price + jump_size * tick_size
spread = spread + jump_size
bid_price = bid_price
\end{verbatim}
whereas in the hypothetical models, it becomes
\begin{verbatim}
simulate D
ask_price = ask_price + jump_size * tick_size * D
spread = spread + jump_size
bid_price = ask_price - spread * tick_size
\end{verbatim}
where D is an independent random variable with the distributions given in Table \ref{Tbl:D}.
\begin{table}[ht!]
	\centering
	\caption{Probability Mass Function of the Direction Random Variable D}
	\label{Tbl:D}
	\begin{tabular}{rrrr}
		\toprule
		Value & LOB 1 & LOB 2 & LOB 3\\ 
		\midrule
		-1 & 0.50 & 0.33 & 0.20 \\
		0 & 0.00 & 0.33  & 0.00 \\
		1 & 0.50 & 0.33 & 0.80 \\ 
		\bottomrule
	\end{tabular}
\end{table}
In other words, the direction of a price movement is independent of the arrival order type; the price can go down with a buy market order or up with a sell market order. When $D=1$, the price moves in the proper direction consistent with the order type, while in the other two cases, the price either does not move or move to the wrong direction. 

Similar logic is applied to orders of type 1-6 and we keep the bid and ask prices unchanged for type 7,8 orders in the alternative LOBs. Finally, we assume the thresholds for the optimal control strategy do not change in the hypothetical LOBs.

\begin{table}[ht!]
\footnotesize
	\centering
	\setlength\tabcolsep{1 pt}	
	\caption{Simulated Backtest on two hypothetical inconsistent LOBs (N=1E6, T=6.5 hours)}
	\label{Tbl:bt3}
	\begin{tabular}{lrrrrrrrr}
		\toprule
		& \multicolumn{2}{c}{Weakly Consistent LOB}  & \multicolumn{2}{c}{Inconsistent LOB 1}  &  \multicolumn{2}{c}{Inconsistent LOB 2} &  \multicolumn{2}{c}{Inconsistent LOB 3} \\ 
	\midrule
	& Unconstrained & Optimal & Unconstrained  & Optimal & Unconstrained  & Optimal & Unconstrained & Optimal \\
	& Trading & Control &  Trading &  Control &  Trading &  Control & Trading & Control \\
	\midrule
	Mean & 8,224 & 6,471 & 12,286 & 10,533 & 12,304 & 10,534  & 9,853 & 8,095 \\
	SD & 10,258 & 516 & 22,830 & 941 & 19,515 & 833 & 16,488 & 716 \\
	Skewness & -0.44 & 0.10 & 0.16 & 0.11 & 0.20 & 0.12 & -0.08 & 0.09\\
	Kurtosis & 7.81 & 3.18 & 7.61 & 3.19 & 7.70 & 3.19 & 7.63 & 3.31 \\
	IR & 0.80 & 12.55 & 0.54 & 11.19 & 0.63 & 12.65 & 0.6 & 11.30 \\
	Overstatement & NA & NA & 49\% & 63\% & 50\% & 63\% & 20\% & 25\% \\ 

	\bottomrule
\end{tabular}
\end{table}

The results in Table \ref{Tbl:bt3} are rather loud and clear: the mean profits are seriously overstated in the three inconsistent LOBs, compared to what would occur in our weakly consistent one. The overstatement of mean profit is around 50-60\% in LOB1 and LOB2. In addition, even the seemingly harmless LOB3, where the direction is right 80\% of the time, still has a nontrivial 20-25\% discrepancy. 

We have stressed throughout this paper that the market maker is often on the wrong side of the trade due to adverse selection. Hence if the price does not move in the appropriate direction consistent with the order type, it can lead to an exaggeration of expected profit and this is most evident in Table \ref{Tbl:bt3}.

Besides, the performance metrics can also be distorted under inconsistent LOBs. For example, under the weakly consistent LOB, the IR improvement by the optimal control is around 16 times but it becomes 20 times under the inconsistent ones; hence it is also possible that a good strategy may appear worse than a poor one under an inconsistent LOB.  Therefore, an optimized trading strategy under an inconsistent framework may not perform well in real-world trading.

\section{Conclusion}

We develop from the ground up a new market-making model which is tailor-made for high-frequency trading under a LOB. In this model, we avoid the common but overly simplistic assumptions of independent price processes, constant volume, one-tick jump and spread, continuous switching without penalty and no market orders. Instead, we build a flexible framework that enforces consistent price movements, allows arbitrary volume, jump, and spread distributions, includes a state-dependent switching cost and permits the use of market orders.

Departing from the classical \citet{Ave2008} framework of regular stochastic control on diffusion, we exploit optimal switching and impulse control on marked point processes. They have proven to be very effective in modeling order-book features such as price-order co-jump, volume, jump size, price-time priority, as well as market orders.

By leveraging the well-known classification of order types in market microstructure and assuming an even distribution of limit orders from a small market maker, the control problem is significantly simplified as we do not need to keep track of the order book shape and the priority of each limit order from the market maker. By further assuming non-stochastic intensities for the arrival processes, the optimal control can be computed numerically by solving an explicit finite-difference scheme for the associated Hamilton-Jacobi-Bellman quasi-variational inequality. Since the scheme is highly parallelizable and an equilibrium is reached very quickly, the computation can be finished within 5 minutes using a 16 cores machine.

The ultimate market-making models may involve full-blown LOB, but currently it is all but impossible to perform optimization on such a high-dimensional object. Our weakly consistent reduced-form model provides a tractable alternative which still includes many of the important order-book features. Our numerical analysis shows that it constitutes a valuable risk-management tool for market makers, greatly reducing the risks associated with providing market liquidity.

Finally, we would like to point out that equation \eqref{HJB} is in fact a system of QVIs (with index $(r_b,r_a)$) involving two nonlocal operators (integral and intervention operators). As far as we know, the theoretical foundations, such as existence, uniqueness, comparison principle, continuity, and linkage with the value function of the stochastic control problem, of the associated viscosity solution have not yet been rigorously established. Moreover, the framework of \citet{Bar1991}, which is commonly used to assert the convergence of numerical schemes to viscosity solutions of PDEs, IPDEs \cite{Bri2004} and QVIs \cite{Azi2018}, needs to be extended, in order to prove the convergence of our numerical scheme \eqref{Eq:scheme}.

\section{Acknowledgments}
The authors would like to thank the participants, in particular Sebastian Jaimungal, in the Stevens High Frequency Finance and Data Analytics Conference 2019, for their comments and suggestions as well as the two anonymous referees for their invaluable feedback.

\begin{appendices}

\section{Brief History of Market-Making Models} \label{appendx}
The early literature on market making appears mostly in the field of market microstructure in finance where researchers study the behavior of various market participants in the financial system. The early models, namely \citet{Gar1976,Ami1980,Sto1978,Ho1981}, are commonly called agent-based models where a monopolistic market maker continuously adjusts his bid and ask quotes in order to control her inventory level. Such models provide a lucid framework to understand the interactions between different market players as well as their impact on the quote-driven market, at the time when the dominant mechanism of security transaction is voice trading and the designated market makers execute majority of the transactions via open outcry.

Another type of market-making models are the pure stochastic models as in \citet{Ave2008,Gue2013,Gui2013}. In those models, the market maker is assumed to be small enough so that she has negligible influence on the order flow. With the rise of electronic LOBs, that allow direct interaction between buyers and sellers, this plausible assumption provide a tractable framework to study market making in the new era.

\subsection{Garman (1976)}
Garman's \cite{Gar1976} model is often regarded as the earliest model of market making, and the title of his paper, market microstructure, develops into a discipline of rigorous study of market mechanism in the field of finance. In Garman's model, there is only one monopolistic market maker and no direct exchange between buyer and seller is allowed. As a result, the market maker has the full price control. However, the rate of incoming Poisson buy and sell order $\lambda_a, \lambda_b$ will depend on the ask and bid price $S_a,S_b$, which he sets at time 0 and the prices will remain the same throughout the whole trading period. At time $0$, he has cash $B_0$ and inventory $Q_0$ and he will go bankrupt when either of them drops to zero. In Garman's setting, the market maker is risk-neutral and he seeks only to maximize the expected profit while avoiding bankruptcy.

Assuming a linear rate function $\lambda_b(s) = \alpha + \beta s, \ \lambda_a(s)=\gamma - \delta s$ with $\gamma > \alpha \ge 0, \ \beta,\delta >0$, in order to avoid running out of inventory or holding infinite amount of stock, the market maker will set the bid and ask prices $S_b, S_a$ such that $\lambda_b(S_b)=\lambda_a(S_a)$. At the same time, she seeks to maximize the profit by solving the static optimization problem
\begin{gather}
\max_{S_b,S_a} (S_a-S_b)(\alpha+\beta S_b) \text{ s.t. }\\
\alpha + \beta S_b = \gamma - \delta S_a
\end{gather}
The solution is $S_b = (\lambda^* - \alpha)/\beta, \ S_a = (\gamma - \lambda^*)/\delta$ where $\lambda^*= (\alpha  \delta + \gamma \beta)/(2 (\beta+\delta))$.

Under Garman's setting, the inventory $Q_t$ is a birth and death process with birth rate $\lambda_{i,i+1} = \lambda_b$ and death rate $\lambda_{i,i-1}=\lambda_a$. From the theory of continuous-time Markov chain, when $\lambda_a = \lambda_b$, the stock ruin probability $\mathbb{P}(Q_t=0 \ \exists t \ge 0 | Q_0=i) = 1 \ \forall i$. In other words, fixing the bid and price at $t=0$ is not viable as the market maker will run out of inventory almost surely.

\subsection{Ho and Stoll (1981)}
Ho and Stoll \cite{Ho1981} extend Garman's model by allowing the bid and ask prices to change continuously over time and they use the stochastic optimal control technique to solve the market-making problem. Same as Garmen, the authors use linear demand/supply functions for the Poisson process of buy and sell orders $N_t^a,N_t^b$, and the dollar value of inventory $I_t$ follows a jump diffusion, where the randomness coming from both maker-making activities and the price fluctuations of the asset. $B_t$, the cash generated from the market-making activities, grows at risk-free rate $r$ while $Y_t$, the wealth of market maker with investment yield $y$, follows a geometric Brownian motion. $S^m$ is a fixed fair price decided by the market maker. Unlike Garmen, the market maker is now risk-averse with a concave utility function $U$. The optimal control problem for market making is stated below:
\begin{gather}
\max_{S_t^b,S_t^a} \E(U(B_T+I_T+Y_T))\\
dB_t = r B_tdt - S_t^b dN_t^b + S_t^a dN_t^a, \ B_0=0\\
dI_t = q I_t dt + S^m (dN_t^b - dN_t^a) + \sigma_I I_t dW_t^I, \ I_0=0\\
dY_t = y Y_tdt + \sigma_Y Y_t dWt^Y, Y_0 > 0 \\
\lambda_t^a = \alpha - \beta (S_t^a - S^m)\\
\lambda_t^b = \alpha - \beta (S^m -S_t^b)
\end{gather}

\subsection{Avellaneda and Stoikov (2008)}
After 27 years, \citet{Ave2008} propose a refinement of \citet{Ho1981}, that forms the foundation of contemporary market-making models. The major difference with \citet{Ho1981} is that instead of modeling a monopolistic market maker, the authors consider a small market maker as price-taker and thus they replace the market-maker's fair price with an exogenous mid price process $S_t^m$, which follows Brownian motion. 

Based on some empirical studies of LOB \cite{Gop2000,Mas2001,Gab2006,Web2005,Pot2003}, Avellaneda and Stoikov suggest to portray the intensity of market order, hitting a limit order at a distance of $d$ (half-spread) from the mid price, in an exponential form $\lambda(d) = A\exp(-k d)$ where $A > 0,k >0$ are constants. Instead of describing the dollar value of inventory $I_t$, the authors use the inventory quantity $Q_t$ (no. of shares). Finally, the interest rate and dividend yield are ignored as the time horizon of a single session of market-making activity is usually just one trading day.
\begin{gather}
\max_{S_t^a,S_t^b} \E(U(B_T+Q_TS_T^m))\\
dB_t = S_t^adN_t^a - S_t^bdN_t^b\\
dQ_t=dN_t^b - dN_t^a\\
dS_t^m=\sigma dW_t\\
\lambda_b(S_t - S_t^b)=A\exp(-k(S_t-S_t^b))\\
\lambda_a(S_t^a - S_t)=A\exp(-k(S_t^a-S_t))
\end{gather}

Although the title of \citeauthor{Ave2008}'s article is called \emph{High-frequency trading in a limit order book}, they do not incorporate any LOB features such as price tick, price-time priority, order size, fee and rebate into their model. The only LOB related claim is that the exponential form of arrival intensity $\lambda$ is coming from empirical studies of LOB.

\subsection{Guilbaud and Pham (2013)}
\citet{Gui2013} introduce a number of LOB features not seen in the AS framework, and the formulation of their control problem is also significantly different. First, the prices of market maker's limit orders $(S_t^a,S_t^b)$ are no longer continuous decision variables, but are either pegged to the best bid/ask\footnote{The best bid and ask prices are still not in the price grid as the mid price is continuous.} or \emph{one tick better}. When the bid-ask spread is only one tick, a one-tick-better limit order means market order. Second, the mid price $S_t^m$ is extended to jump diffusion (L\'{e}vy process)\footnote{$\xi$ in equation \eqref{levy} is the continuous jump size of L\'{e}vy process.} and the discrete bid-ask spread $S_t$ is modeled by an independent continuous-time Markov chain. Besides, market maker can choose the size of limit orders $L_t^a,L_t^b$ as well as the time $\tau_n$ and size $\zeta_n$ of market orders, which are used to remove excess inventory. Lastly, the final liquidation value includes the cost of crossing the spread $(|Q_T|S_T/2)$ and a non-proportional exchange fee $\eta$.
\begin{gather}
\max_{S_t^a,S_t^b,L_t^a,L_t^b,\tau_n,\zeta_n} \E \Big( U \big( B_T+Q_TS_T^m -|Q_T|S_T/2-\eta \big) \Big)\\
B_t = \int_0^t S_s^a L_s^a dN_s^a - \int_0^t S_s^b L_s^a dN_s^b - \sum_{\tau_n \le t} \big( \zeta_n S_{\tau_n}^m + |\zeta_n| S_{\tau_n}/2 + \eta \big)  \\
Q_t= \int_0^t L_s^b dN_s^b - \int_0^t L_s^a dN_s^a + \sum_{\tau_n \le t} \zeta_n\\
S_t^m= \mu t + \sigma W_t + \int_{(0,t] \times (-1,1)} \xi \widetilde{N}^m(ds \times d\xi) + \int_{(0,t] \times (-1,1)^c} \xi N^m(ds \times d\xi) \label{levy} \\
S_t \sim \text{ independent continuous-time Markov chain }
\end{gather}
where $\mu, \sigma > 0$ are constants.

\end{appendices}

\bibliographystyle{unsrtnat2}
\bibliography{output}

\end{document}